\documentclass[traditabstract]{aa}
\usepackage{graphicx}
\usepackage{natbib}
\usepackage{txfonts}

\usepackage{booktabs}
\bibpunct{(}{)}{;}{a}{}{,}

\usepackage{subfig}
\usepackage{numprint}
\npthousandsep{,}
\usepackage{hyperref} 
\hypersetup{
  colorlinks=true,
  pdfborder=2 2 0.,
  linkcolor=blue,
  anchorcolor=black,
  citecolor=blue,
  urlcolor=black,
}

\begin{document}

  \title{The age of the Milky Way inner stellar spheroid from RR Lyrae population synthesis}
  
  \author{A. Savino\inst{1}, A. Koch\inst{1}, Z. Prudil\inst{1}, A. Kunder\inst{2} \and R. Smolec\inst{3}}

\institute{Astronomisches Rechen-Institut, Zentrum f\"{u}r Astronomie der Universit\"{a}t Heidelberg, M\"{o}nchhofstr. 12-14, D-69120 Heidelberg, Germany \email{A.savino@uni-heidelberg.de} 
           \and Saint Martin\textsc{\char13}s University, 5000 Abbey Way SE, Lacey, WA, 98503
           \and Nicolaus Copernicus Astronomical Center, Polish Academy of Sciences, ul. Bartycka 18, 00-716 Warszawa, Poland
           }

 \abstract{
 The central kiloparsecs of the Milky Way are known to host an old, spheroidal stellar population, whose spatial and kinematical properties set it apart from the boxy/peanut structure that constitutes most of the central stellar mass. The nature of this spheroidal population, whether a small classical bulge, the innermost stellar halo or a population of disk stars with large initial velocity dispersion, remains unclear. This structure is also a promising candidate to host some of the oldest stars in the Galaxy.  Here we address the topic of the inner stellar spheroid age, using spectroscopic and photometric metallicities for a sample of 935 RR Lyrae stars that are constituents of this component. By means of stellar population synthesis, we derive an age-metallicity relation for RR Lyrae populations.We infer, for the RR Lyrae stars in the bulge spheroid, an extremely ancient age of $13.41 \pm 0.54$ Gyr and conclude they were among the first stars to form in what is now the Milky Way galaxy. Our age estimate for the central spheroid shows remarkable agreement with the age profile that has been inferred for the Milky Way stellar halo, suggesting a connection between the two structures. However, we find mild evidence for a transition in the halo properties at $r_{\rm GC} \sim 5$~kpc. We also investigate formation scenarios for metal-rich RR Lyrae stars, such as binarity and helium variations, and whether they can provide alternative explanations for the properties of our sample. We conclude that, within our framework, the only viable alternative is to have younger, slightly helium-rich, RR Lyrae stars, a hypothesis that would open intriguing questions for the formation of the inner stellar spheroid.

}

\authorrunning{A. Savino et al.}
\titlerunning{The age of RR Lyrae stars in the Galactic bulge}
  \maketitle

%_____________________________________________________________________

\section{Introduction}

Among the most challenging regions of the Milky Way to observe, the Galactic bulge has been subject to a tireless investigation, since its discovery. These endeavours slowly unveiled it as a region full of chemodynamical complexities \citep[e.g.][and references therein]{Rich13, McWilliam16, Barbuy18a}. The presence of a stellar bar in our galaxy was firmly established towards the end of the last century \citep[e.g.][]{Binney91, Blitz91, Stanek94}. It later became clear that most of the stellar mass in the Milky Way central regions belongs to a boxy/peanut structure, with its characteristic X shape and cylindrical rotation \citep[e.g.][]{Howard09,McWilliam10, Nataf10, Kunder12, Wegg13}. The favoured explanation for the origin of these features invokes assembly through dynamical instabilities of the inner thin/thick disks \citep[e.g.][]{Kormendy04, Shen10}. 

However, there is solid evidence pointing at the bulge to be best described as a superposition of multiple stellar components. This becomes clear if we shift the focus from stars with [Fe/H]$\, \approx 0$, which dominate the mass budget, towards lower metallicity stars, for which the morphological and kinematical signature of the boxy/peanut structure progressively disappears, until one is left with a spheroidal, concentrated, pressure-supported structure \citep[e.g.][]{Dekany13, Kunder16, Arentsen20,Kunder20}.  Among the possible origins for this stellar population, a small classical bulge or  the innermost extension of the Milky Way halo have been suggested \citep[e.g.][]{Minniti99,Shen10,Perez-villegas17,Rojas-Arriagada17}. It may also be that the multiple stellar components originated together in one disk, and the different initial velocity dispersions and ages are manifest differently in today's bulge \citep{Debattista17}. The nature of this stellar structure is yet to be unambiguously determined and additional characterisation is needed to distinguish among the possible scenarios.

Additionally, the metal-poor ([Fe/H]$\, \lesssim -1.0$) stellar population of the central Milky Way spheroid is of great interest from a broader cosmological perspective.The current cold dark-matter paradigm predicts dark-matter haloes to grow in an inside-out fashion. This is reflected in the formation of the early stellar spheroid and the inner regions of the galactic potential well have been suggested to host the oldest living stellar relics of galaxy formation \citep[e.g.][]{White00, Brook07,Tumlinson10}. A natural consequence of this scenario is that the oldest stars in the Milky Way bulge are the best candidates to carry a pristine chemical imprint of the first, metal-free stars \citep[e.g.][]{Tumlinson10, Chiappini11, Koch16}.

Evidence that stars in the Milky Way halo tend to be older at small Galactocentric radii has been recently obtained through chronography of blue horizontal branch (HB) stars \citep{Santucci15, Carollo16, Das16}. However, these studies focused on stars further than 5~kpc from the Galactic centre, leaving the inner stellar spheroid unexplored. The ancient nature of this central stellar population is effectively revealed by the presence of a sizeable population of RR Lyrae (RRL) stars, in the Milky Way bulge, with high velocity dispersion and little or no rotational signature \citep{Kunder16}. Because RRLs originate from low-mass stars and are currently on the HB populating the instability strip (IS), they are often quoted to have ages greater than 10-11~Gyr \citep{Walker89,Marconi15}. While already of great value for interpreting the archaeological record of local stellar populations, this figure spans a range of possible ages of roughly 3.5~Gyr or 25\% of the age of the Universe. However, RRLs have the potential to be much more powerful age tracers. Due to the stringent effective temperature constraints required to trigger stellar pulsation, the metallicity at which HB stars manifest as RRLs is dependent on the stellar population age, as these are two of the main parameters governing the temperature range of HB stars \citep[e.g.][]{Gratton10}. This was already noted by \citet{Lee92}, who suggested a very old age for the RRLs in the bulge. This argument, however, was based on a comparative analysis based on halo RRLs, due to the large uncertainties of HB population models of the time.

In this paper, we present a theoretical framework that allows us, for the first time, to derive a precise absolute age scale for populations of RRLs, based on their observed metallicity distribution. We combine this with a spectroscopic analysis of a large sample of bulge RRLs, showing that stars in the inner Galactic spheroid formed at an extremely ancient time, prior to any known stellar population in the Milky Way. The paper is organised as follows. In Sect.~\ref{Obs} we describe the spectroscopic dataset of bulge RRLs and the procedure employed to measure their metallicity. In Sect.~\ref{Model} we describe our stellar population synthesis approach, derive an age-metallicity relation for RRL populations and the implications for the bulge stellar spheroid. In Sect.~\ref{Alternatives} we explore alternative formation scenarios for high-metallicity RRLs. In Sect.~\ref{Conclusions} we summarise our results and discuss its broader implications.

%----------------------------------------------------------------------------------------------------------------------------
%----------------------------------------------------------------------------------------------------------------------------
%--------------------------------------      SECTION 2      ------------------------------------------------------------
%----------------------------------------------------------------------------------------------------------------------------
%----------------------------------------------------------------------------------------------------------------------------

\section{The metallicity of RR Lyrae stars in the bulge}
\label{Obs}
As RRL metallicities will be our primary instrument of investigation, we devote the first step of our analysis to the characterisation of the metallicity\footnote{In this paper the term metallicity is used to refer to the [Fe/H] abundance, even when this quantity is inferred indirectly, e.g. through modelling of calcium lines.} of RRLs in the central regions of the Milky Way. Currently, spectroscopic metallicities for RRLs in the inner Galaxy are only available for limited samples, with the largest one being the 59 stars observed in Baade's Window by \citet{Walker91}, who inferred an average [Fe/H]~$\sim -1.0$. Our view of the RRLs in the bulge improved significantly with the advent of the large variability surveys, such as MACHO, OGLE and VVV \citep{Alcock00,Udalski02,Minniti10}. As the chemistry of RRLs leaves an imprint on their light curve, these datasets allows one to derive photometric metallicities for hundreds to tens of thousands of objects, reporting mean [Fe/H] values between -1 and -1.3 \citep[e.g.][]{Kunder08,Pietrukowicz15,Minniti17,Dekany18}.

In spite of the large spectroscopic coverage that the bulge has been subject to in the last decade, as of today no spectroscopic metallicities have been determined for a large sample of RRLs in this region. The reason is that the vast majority of the spectroscopic surveys targeting the bulge have selection criteria that are tuned  for much cooler stars, such a red giants \citep{Howard08} and red clump stars \citep{Freeman13}. In the following, we describe our derivation of spectroscopic metallicities, using data from the only bulge spectroscopic program that targeted a large population of RRLs.

%----------------------------------------------------------------------------------------------------------------------------
%--------------------------------------      DATA     ------------------------------------------------------------
%----------------------------------------------------------------------------------------------------------------------------

\subsection{Data}
We use spectra from the data release 1 of the BRAVA-RR survey \citep[][AAT PropIDs: 2013A-05; PI: D. Nataf; NOAO PropIDs: 2014A-0143 and 2015B-071; PI: A. Kunder]{Kunder16}, which has been carried out with the AAOmega multi-fiber spectrograph on the Anglo-Australian Telescope, from May 2013 to August 2015. Our dataset consists of multi-epoch observations of 945 type \textit{ab} RRLs\footnote{Although RRLs are classified in \textit{ab}, \textit{c} and \textit{d} subtypes, our analysis and the majority of the referenced works refer specifically to the \textit{ab} RRLs, the most common type of RRLs and those that pulsate in the fundamental mode. In the following, we will use the term RRL with the implicit reference to the \textit{ab} pulsators only.}, totalling 4002 spectra taken over the entire range of pulsation phase. The stars are located in 4 fields, in the southern Galactic plane, roughly $2^\circ$ wide and centred at $l$, $b$ = ($-3$,$-5$), ($-1$,$-4$), ($2$,$-3$) and ($3$,$-5$). The stars have heliocentric distances, from \citet{Kunder20}, between 5.0 and 16.5~kpc, with roughly 90\% of the sample between 7.0 and 11.5~kpc. The approximate effective-temperature range of our targets is 5500-8000~K \citep{Jurcsik18}.

\begin{figure}
        \subfloat[][]
	{\includegraphics[width=0.5\textwidth]{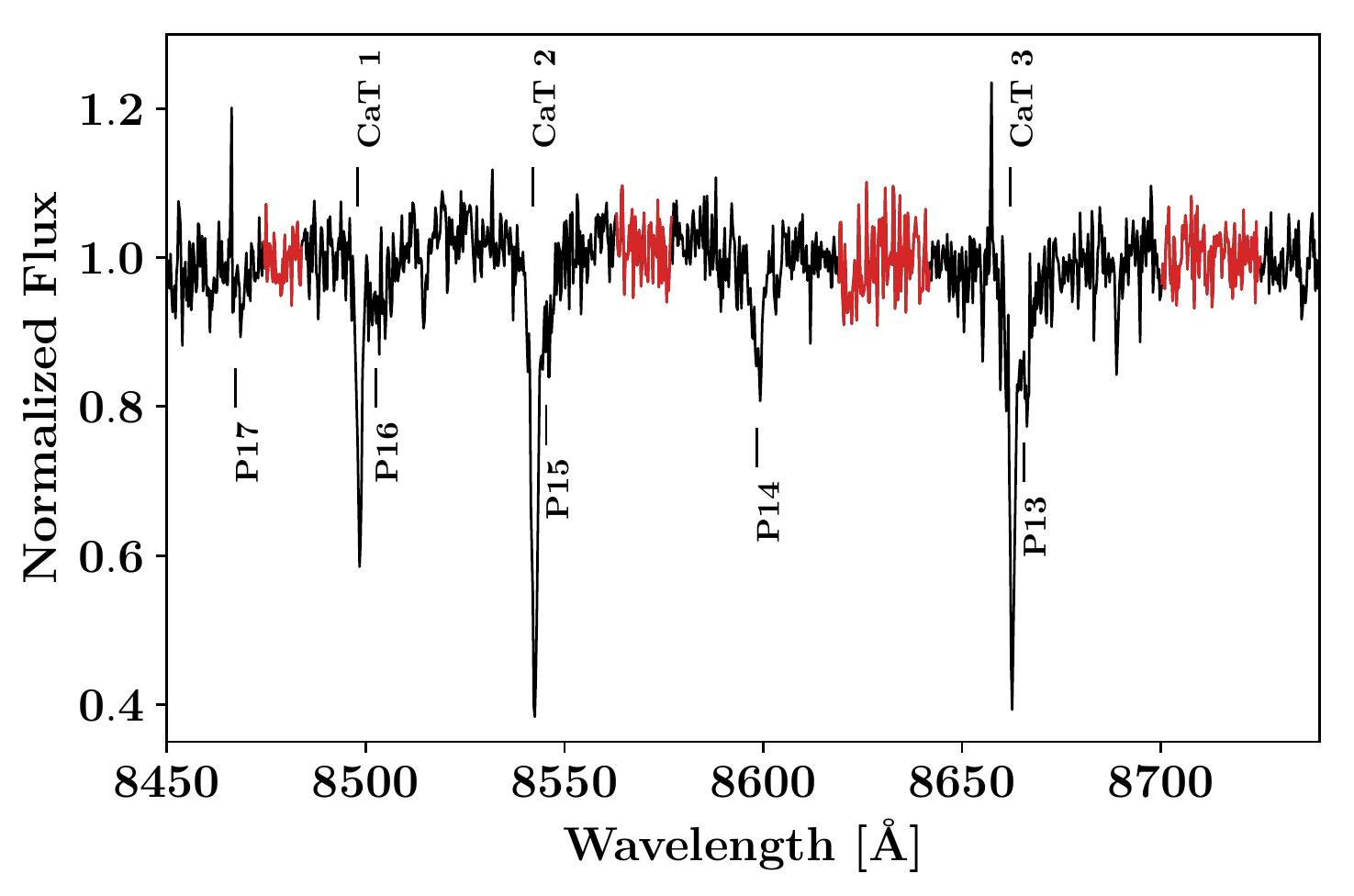} \label{fig:spec}} \quad
	 \subfloat[][]
	{\includegraphics[width=0.5\textwidth]{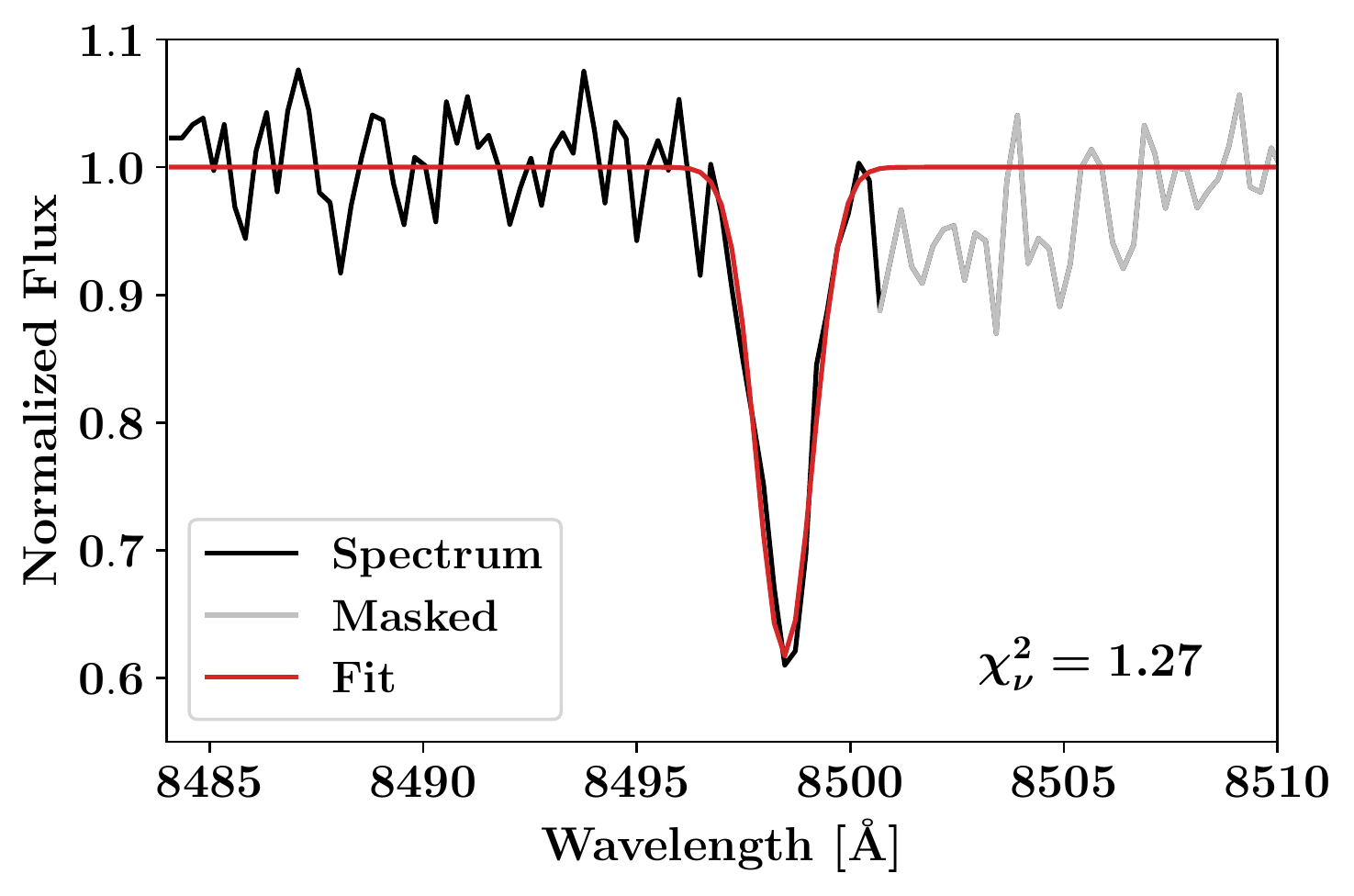} \label{fig:line} } \quad
  \caption{a) Representative, continuum normalised, spectrum from our dataset. The red regions are those used for the continuum fitting. The absorption features of the calcium triplet and the hydrogen Paschen series are indicated. b) Zoom on the calcium triplet 8498~$\AA$ line for the same spectrum. The red line shows the best-fit profile. The grey region, containing the Paschen P16 line, was excluded from the fit.}
\end{figure}

The spectra are centred on the calcium II triplet (CaT) region, covering a wavelength range of about 8300-8800 $\AA$, with a resolution of $R\sim \numprint{10000}$. The individual spectra cover a range of signal-to-noise (S/N) ratios, up to 80~px$^{-1}$. The average S/N is 19~px$^{-1}$, with standard deviation of 10. As virtually every star in our sample has multiple observations, the measurements can be statistically combined, and the equivalent S/N per object has an average of 40 px$^{-1}$ and standard deviation of 18. Figure~\ref{fig:spec} shows the CaT region for a representative spectrum. The regions highlighted in red are the ones used for continuum normalisation, as described in the next section. From this initial sample, we removed all spectra observed near maximum light, as justified in Sect.~\ref{specmets}, ending up with a final working sample consisted of 3050 spectra for 935 stars. Of these stars, 818  have multiple spectra, ranging in number from 2 to 15. The median number of spectra per star is 3.

%----------------------------------------------------------------------------------------------------------------------------
%--------------------------------------      METS     ------------------------------------------------------------
%----------------------------------------------------------------------------------------------------------------------------

\subsection{Metallicity derivation}
\label{specmets}
The CaT feature is extensively exploited to derive chemical and kinematical properties of large samples of stars. It is a strong triplet, easily observable even down to low resolutions and low S/N, and it is thus very well suited for radial velocity measurements, a primary reason for its selection for the BRAVA-RR. The CaT also carries information about metallicity and it has been routinely used to measure iron abundance when sample size, target faintness, or metal-poor stars that lack strong spectral features make high-resolution spectroscopy challenging \citep[e.g.][]{Battaglia06,Koch06,Dacosta16,Matijevic17}. However, to extract metallicity information from the CaT, a calibration is required between the strength of this feature and [Fe/H]. This is a challenge for our analysis, as the vast majority of such relations, empirical or theoretical, have been derived for the cool atmospheres of red giants above the HB \citep[e.g.][]{Armandroff91,Rutledge97,Cole04, Battaglia08,Starkenburg10} and are therefore not applicable to our RRL sample, which has significantly warmer atmospheres.

So far, the only available calibration for our purpose is that of \citet[hereafter W12]{Wallerstein12}, who used 30 RRLs in the Solar vicinity, with $\rm -2.4\leq[Fe/H]\leq-0.15$, to derive an empirical relation between [Fe/H] and the equivalent width (EW) of the CaT 8498~$\AA$ line alone (in contrast to the red giant calibrations that use up to all three triplet lines). Only the bluest CaT line is used here because in the warm atmospheres of RRLs the hydrogen Paschen series becomes manifest, progressively increasing in strength with increasing effective temperature. Two of these lines (P13 and P15) blend severely with the CaT lines at 8542 and 8662~$\AA$, rendering their modelling problematic. The 8498~$\AA$ line also lies very close to the Paschen P16 feature, but the two lines are generally separable at this resolution. Therefore, for the rest of this analysis, we focused on this CaT line and derived the metallicities with the W12 calibration:
\begin{equation}
\rm [Fe/H] = -3.846(\pm0.155)+EW\cdot0.004(\pm0.0002).
\label{eq:specmets}
\end{equation}

After shifting the spectra to the rest-frame wavelengths, using the radial velocities from \citet{Kunder16}, we model the continuum with a second order polynomial fit, with iterative three-sigma clipping, to the continuum bandpasses defined in \citet{Cenarro01}. These regions of the spectrum are reliable continuum tracers over a wide range of spectral types, a very desirable feature as RRLs can exhibit up to $\sim2000$~K variations in effective temperature during their pulsation cycle \citep{Pancino15,Jurcsik18}. The CaT EW is measured on the normalised spectrum by fitting a Gaussian profile. Although a superposition of a Gaussian and a Lorentzian profiles has been shown to provide a more accurate EW measurement \citep{Cole04}, the deviation from a purely Gaussian profile is expected to be relevant only for the most metal-rich stars in our sample. More importantly, the calibration by W12 makes use of EWs measured with a Gaussian profile, therefore using the same procedure ensures consistency. We exclude from the fit wavelengths redder than 8500~$\AA$, to avoid contamination from the P16 line. An example of a typical line fit is provided in Fig.~\ref{fig:line}. The reduced chi-square distribution of our line fits has a median of 1.77 and spans a range of 1.15 to 3.26 from the 15.87th to the 84.13th percentile.

\begin{figure}
\centering
\includegraphics[width=0.5\textwidth]{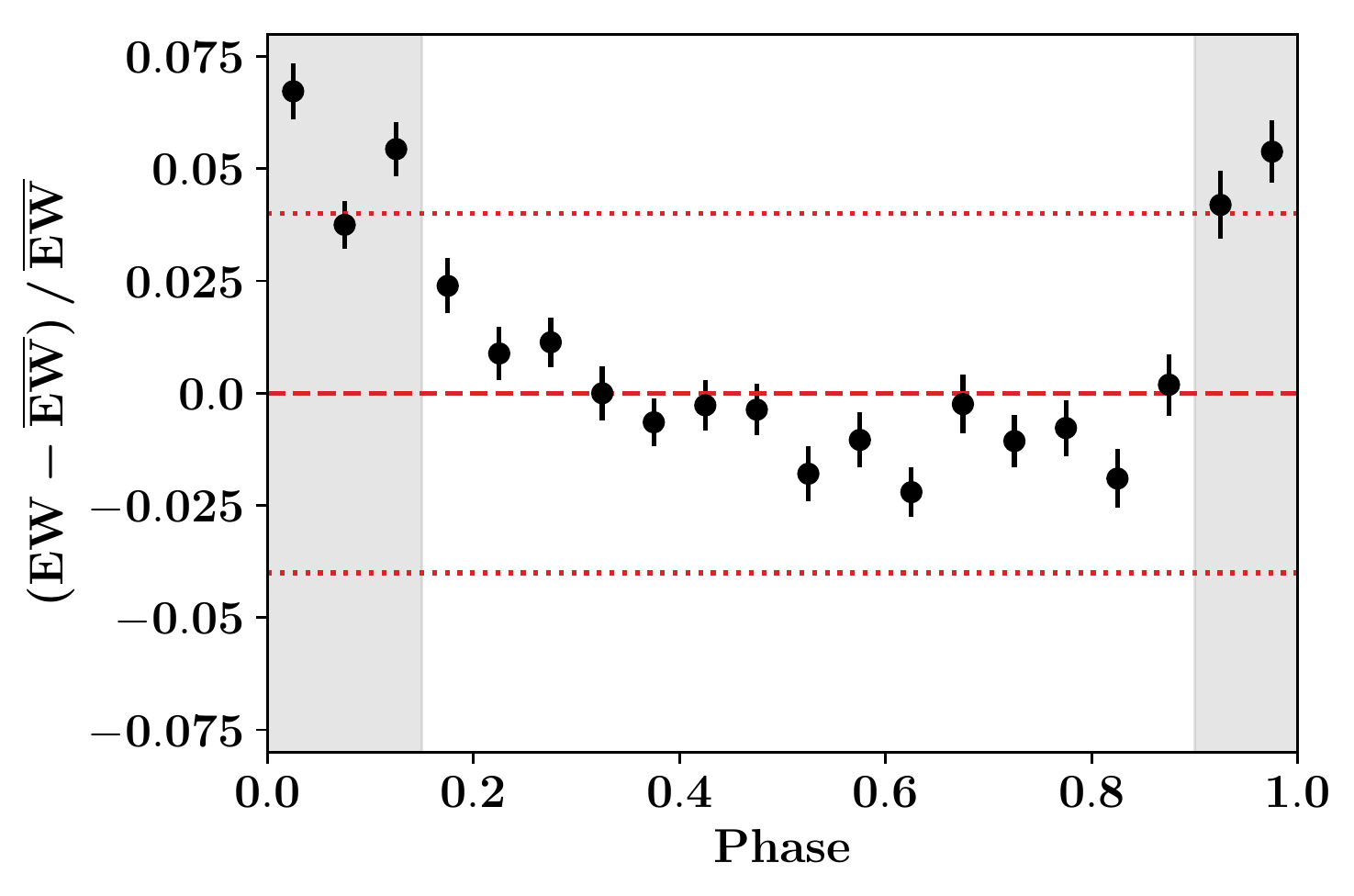}
\caption{ Median fractional variation of the calcium triplet 8498~\AA\, equivalent width as function of pulsation phase. The dashed red line marks the zero. The dotted red lines mark the average uncertainty in the mean equivalent widths used to derive the metallicity. The grey shaded regions show the phase ranges excluded from our analysis.}
\label{fig:Phases}
\end{figure}

\begin{figure*}
       \subfloat[][]
	{\includegraphics[width=0.5\textwidth]{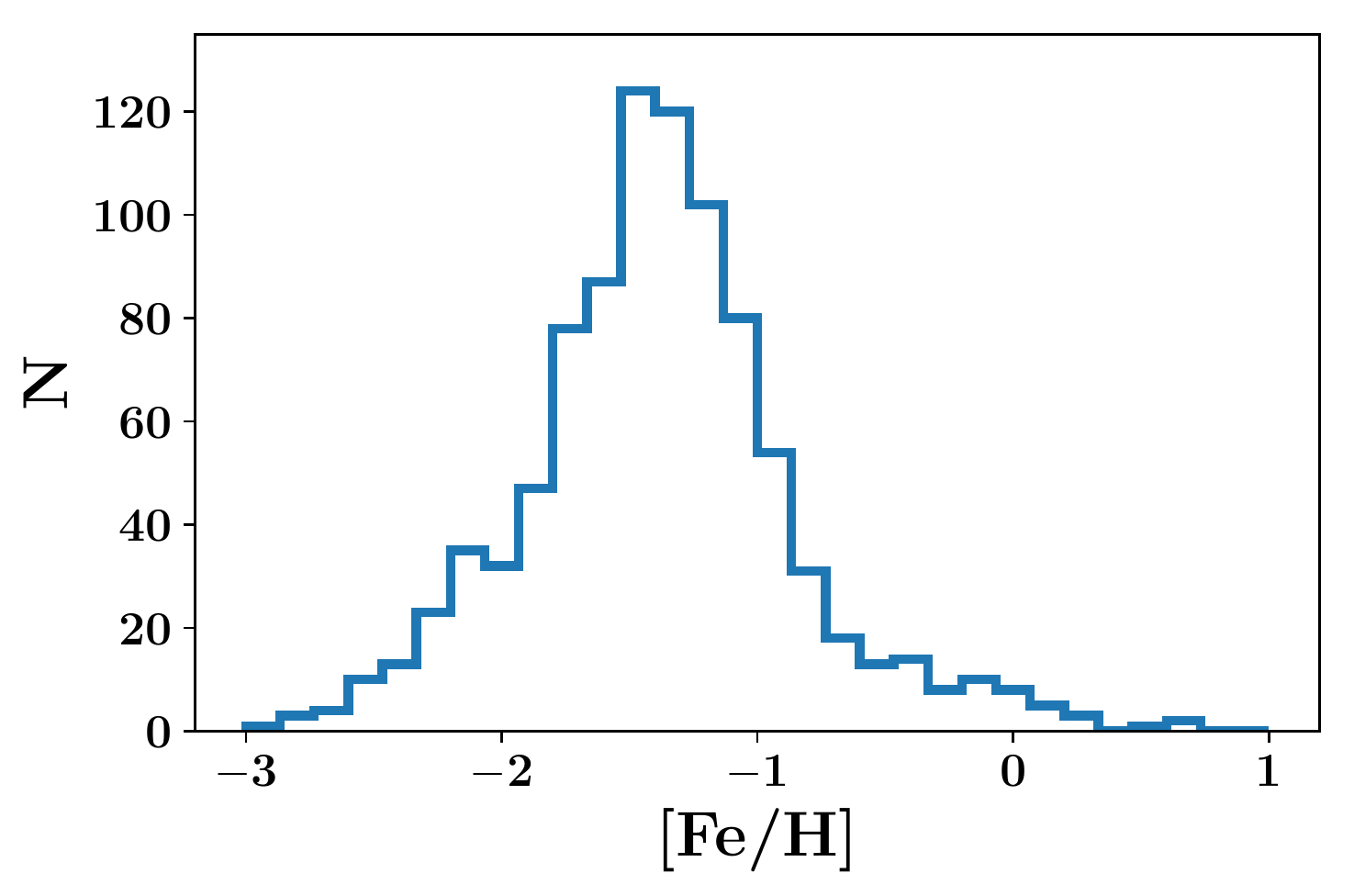} \label{fig:MDF}} \quad
	 \subfloat[][]
	{\includegraphics[width=0.5\textwidth]{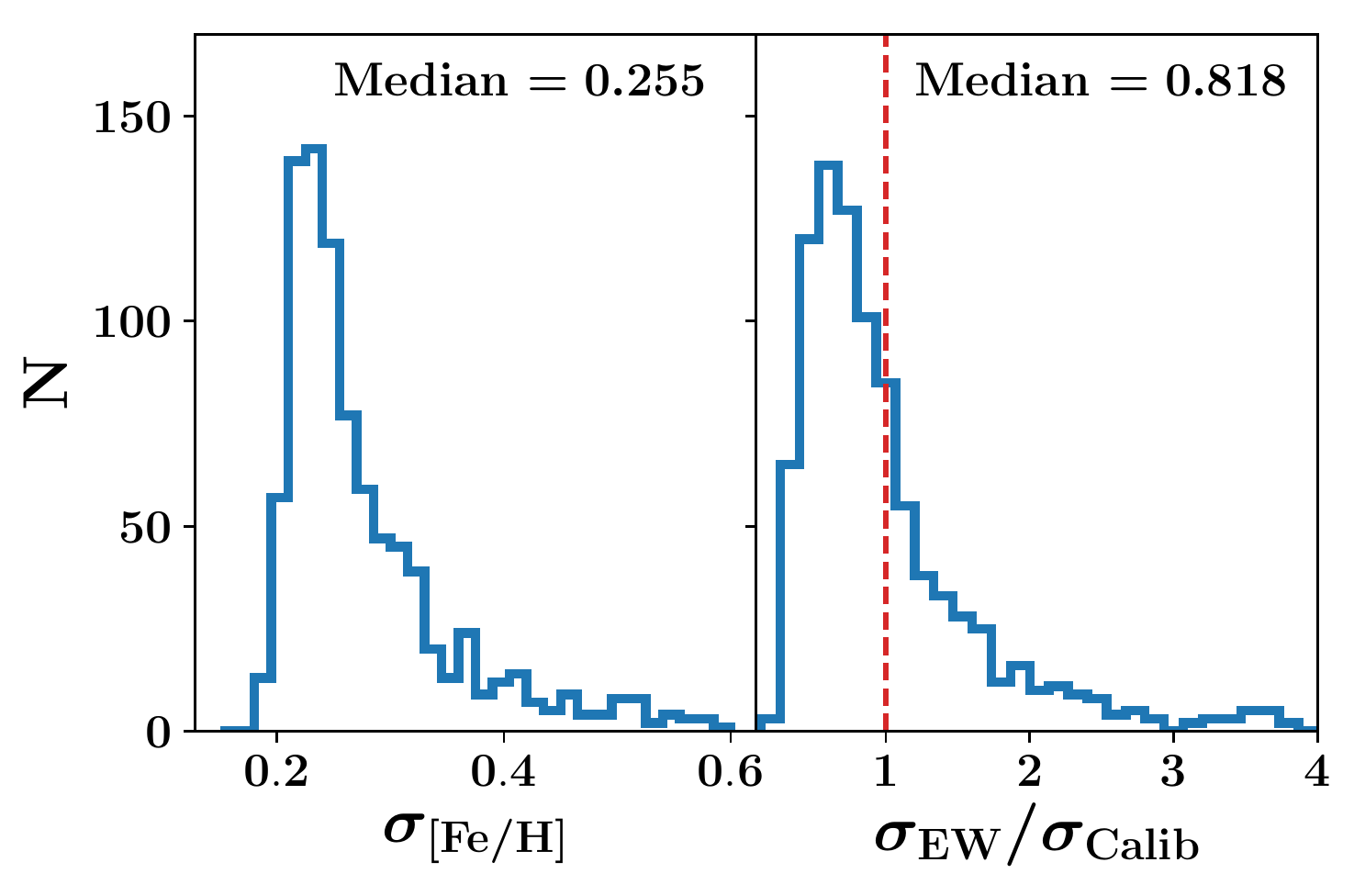} \label{fig:MDF_unc}} \quad
   \caption{a) Observed metallicity distribution function for our sample of RR Lyrae stars. b) Left panel: distribution of metallicity uncertainties. Right panel: distribution of $\sigma_{\rm EW}/\sigma_{\rm Calib}$, showing the dominant contribution to the metallicity uncertainties. The red dashed line marks unity.}
    \label{fig:rad}
\end{figure*}

Errors on the EWs were derived by means of a Monte-Carlo approach. For each spectrum, we generated a sample of 50 independent continuum normalisation polynomials, sampling the uncertainties on the continuum fit parameters. For each of these 50 normalisation, we generated 50 independent Gaussian profiles, accordingly to the line fit uncertainties. The standard deviation on the EW of these 2500 profiles was used as uncertainty on our measurement. As already mentioned, most of our stars we have multiple exposures, ranging in number from 2 to 15. However, as these observations are often taken over a wide range of pulsation phases, it is not wise to stack them and obtain a higher S/N spectrum. Instead, we measure the EWs on each individual spectrum and then combine the measurements statistically, obtaining a weighted average EW and its correspondent uncertainty, that are then used in the W12 relation to obtain the metallicity value.

We note that Eq.~\ref{eq:specmets} does not include any term containing the pulsation phase of the RRL. W12 conclude, from the continuous observations of a single star, that the EW variation along the pulsation cycle is less than 10\%. We followed up on this issue by calculating the fractional difference between the EW obtained from each individual spectrum and the average EW of the RRL it belongs to. The median variation as function of pulsation phase\footnote{Pulsation phases in this paper are defined so that phase 0/1 corresponds to maximum light.} is reported in Fig.~\ref{fig:Phases}. Along most of the pulsation cycle, the CaT EW remains fairly stable, with variations of the order of 2\%. For comparison, the median error on the average EWs (dotted red line in Fig.~\ref{fig:Phases}) is 4\% and, as it is shown below, it is itself a subdominant source of error compared to achievable precision allowed by the calibration uncertainties. The CaT EW becomes mildly overestimated when measured at phases between 0 and 0.15, and between 0.9 and 1. These pulsation stages, close to maximum light, are associated with strong shocks in the stellar atmosphere and are usually deemed unsuitable for spectroscopic determinations \citep{Clementini95, For11, Wallerstein12, Pancino15}. For this reason, we exclude them from our analysis and calculate metallicities only with spectra taken at phases between 0.15 and 0.9.

The measured metallicity distribution function (MDF) for our sample is shown in Fig.~\ref{fig:MDF}. Overall, the MDF covers a wide range of metallicities, with a single prominent peak. The median metallicity for this sample is $[{\rm Fe/H}]_{\rm Med} = -1.39$. The uncertainty distribution of our metallicity measurements is reported in the left panel of Fig.~\ref{fig:MDF_unc} and shows that the great majority of our errors lie between 0.2 and 0.3 dex, with a mild positive correlation with metallicity. The uncertainty distribution almost entirely accounts for the width of the observed MDF peak, suggesting that the true underlying MDF has very narrow, dominant, component, superimposed to wide metal-rich and metal-poor tails. This was already believed to be the case from previous photometric metallicity investigations \citep[e.g.][]{Pietrukowicz15, Dekany18}. The shape of the MDF also justifies our choice, throughout this paper, to use the median [Fe/H] to characterise the metallicity distributions, as it is less sensitive to the wide tails of the distribution and traces more closely the main RRL population (see Sect.~\ref{Alternatives} for a discussion about the nature of the metal-poor and metal-rich tails).

The errors on the metallicity can be represented as the quadratic sum of two components, one coming from the EW uncertainties and one coming from the calibration coefficient uncertainties, so that $\sigma_{{\rm[Fe/H]}}=\sqrt{\sigma_{\rm EW}^2+\sigma_{\rm Calib}^2}$. The ratio of these two components is shown in the right panel of Fig.~\ref{fig:MDF_unc}. It is evident that, for the majority of stars in our sample, the W12 calibration is the dominant source of uncertainty in our metallicity determination. We stress that the work of W12, while an important first step to calibrate a CaT metallicity relation in warm stars, is based on a modest sample of stars. We feel that a much more extensive analysis is needed to investigate the detailed behaviour of the CaT strength in the complicated, dynamic, atmosphere of RRL pulsators. Quantitatively characterising the contamination from the Paschen lines would also be valuable, as it would allow to exploit all of three CaT lines, increasing the measurement robustness.

%----------------------------------------------------------------------------------------------------------------------------
%--------------------------------------      PHOTOMETRIC METS      --------------------------------------------
%----------------------------------------------------------------------------------------------------------------------------

\subsection{Comparison with photometric metallicities}
\begin{figure}
\centering
\includegraphics[width=0.5\textwidth]{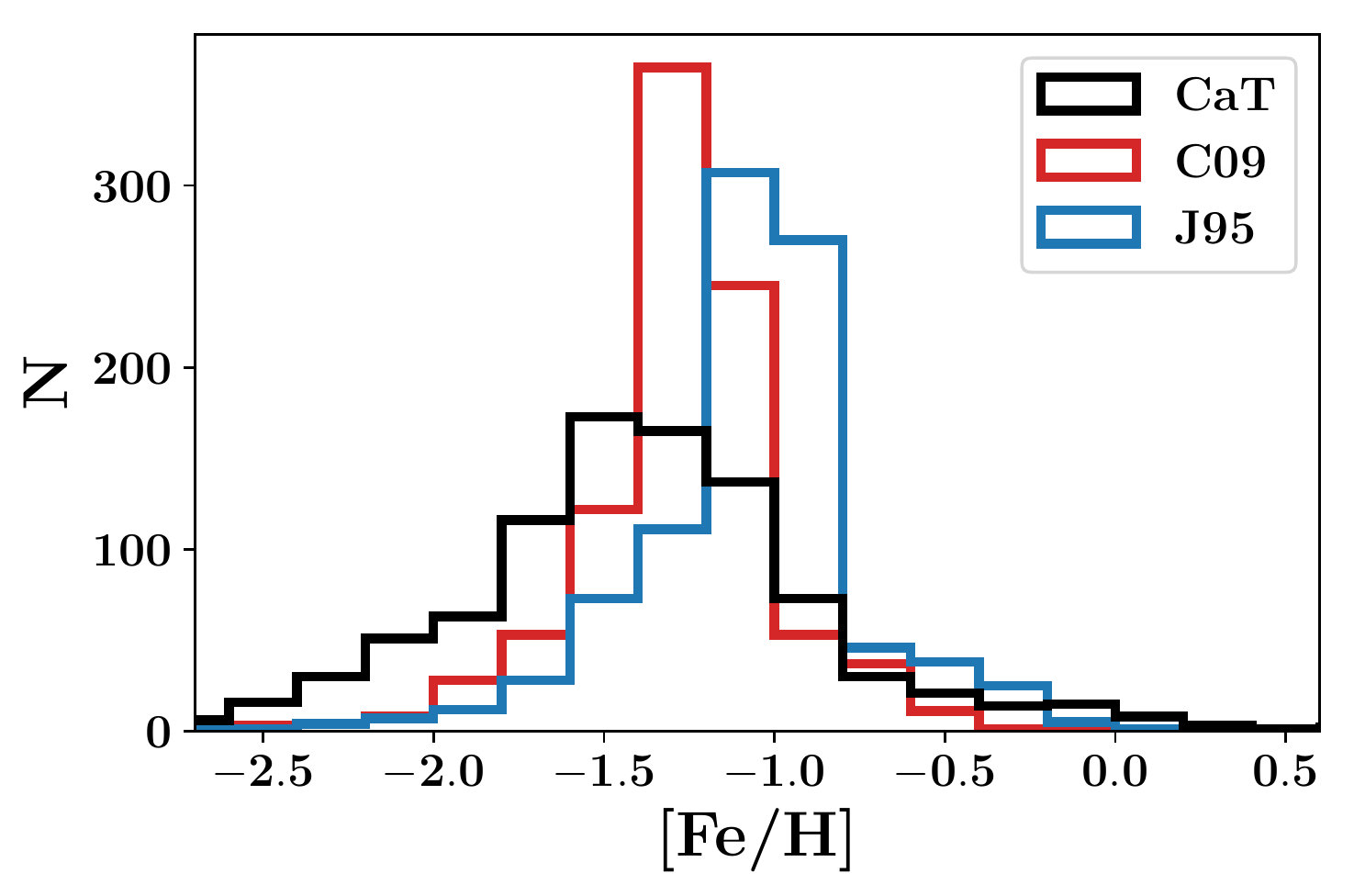}
\caption{ Spectroscopic metallicity distribution obtained from our calcium triplet modelling (black histogram), superimposed to the photometric metallicity distribution, both in the original \citet{Jurcsik95} metallicity scale (blue) and transposed to the \citet{Carretta09b} scale (red).}
\label{fig:Photmet}
\end{figure}

As mentioned already, the shape of the light curve of an RRL carries information about the star's chemical composition. This means that, with sufficient variability coverage, fairly accurate photometric metallicities can be obtained for these objects \citep[e.g.][]{Kovacs95, Smolec05,Hajdu18}. Therefore, it is an instructive exercise to compare spectroscopic and photometric determinations for our sample (as illustrated in Fig.~\ref{fig:Photmet}), so that we can quantify the degree of confidence for the metallicity estimate of this stellar population.

The recipes to derive the iron abundance from the variability information of RRLs are numerous and depend in part on the type of variability data available. As all stars in our sample have extensive variability coverage from the OGLE III and IV data releases, the most direct way to proceed is to use the formula introduced by \citet[hereafter S05]{Smolec05}:
\begin{equation}
[{\rm Fe/H}]_{\rm S05} = -6.125 -4.795\cdot P +1.181\cdot \phi_{31} + 7.876\cdot A_2,
\end{equation}
where $P$ is the star's pulsation period, and $\phi_{31}$ and $A_2$ are parameters derived from the Fourier decomposition of the $I$ band light curve (see S05 for more details). The formula has been calibrated in the metallicity range $\rm -1.7 \leq [Fe/H] \leq 0.1$. The MDF obtained through this relation is the blue histogram in Fig.~\ref{fig:Photmet}. We can see that photometric metallicities obtained in this way tend to be fairly more metal rich than our CaT-based determination. Specifically, the median metallicity of this MDF is $[{\rm Fe/H}]_{\rm Med} = -1.03$.

It is worth mentioning that any [Fe/H] value not directly inferred from the modelling of iron lines has to be anchored to an empirical metallicity scale. Depending on the scale adopted, non-negligible differences can arise in the metallicity inferred for a given star. The formula provided by S05 produces metallicities that are on the \citet[hereafter J95]{Jurcsik95} scale. Among other metallicity scales commonly adopted, it is of particular relevance that derived by \citet[hereafter C09]{Carretta09b}, using thousands of high-resolution spectra in Galactic globular clusters. As globular clusters have long been among the primary calibrators for low-mass stellar evolution models, one could argue that metallicities measured on the C09 scale provide the fairest element of comparison with theoretical stellar population models. We compared the metallicity values obtained by \citet{Jurcsik96} for a sample of RRLs in four Galactic globular clusters\footnote{The \citet{Jurcsik96} sample comprises six Galactic globular clusters but we used only the clusters with $[{\rm Fe/H}]_{\rm J95} > -1.6$, as photometric metallicities on the J95 scale have been shown to be systematically overestimated in the low-metallicity regime.} with the cluster metallicities provided by C09 and, by means of linear regression, we obtained the following conversion:

\begin{equation}
[{\rm Fe/H}]_{\rm C09} = [{\rm Fe/H}]_{\rm J95}\cdot0.833(\pm 0.048)-0.378 (\pm0.052).
\end{equation}

\begin{figure}
\centering
\includegraphics[width=0.5\textwidth]{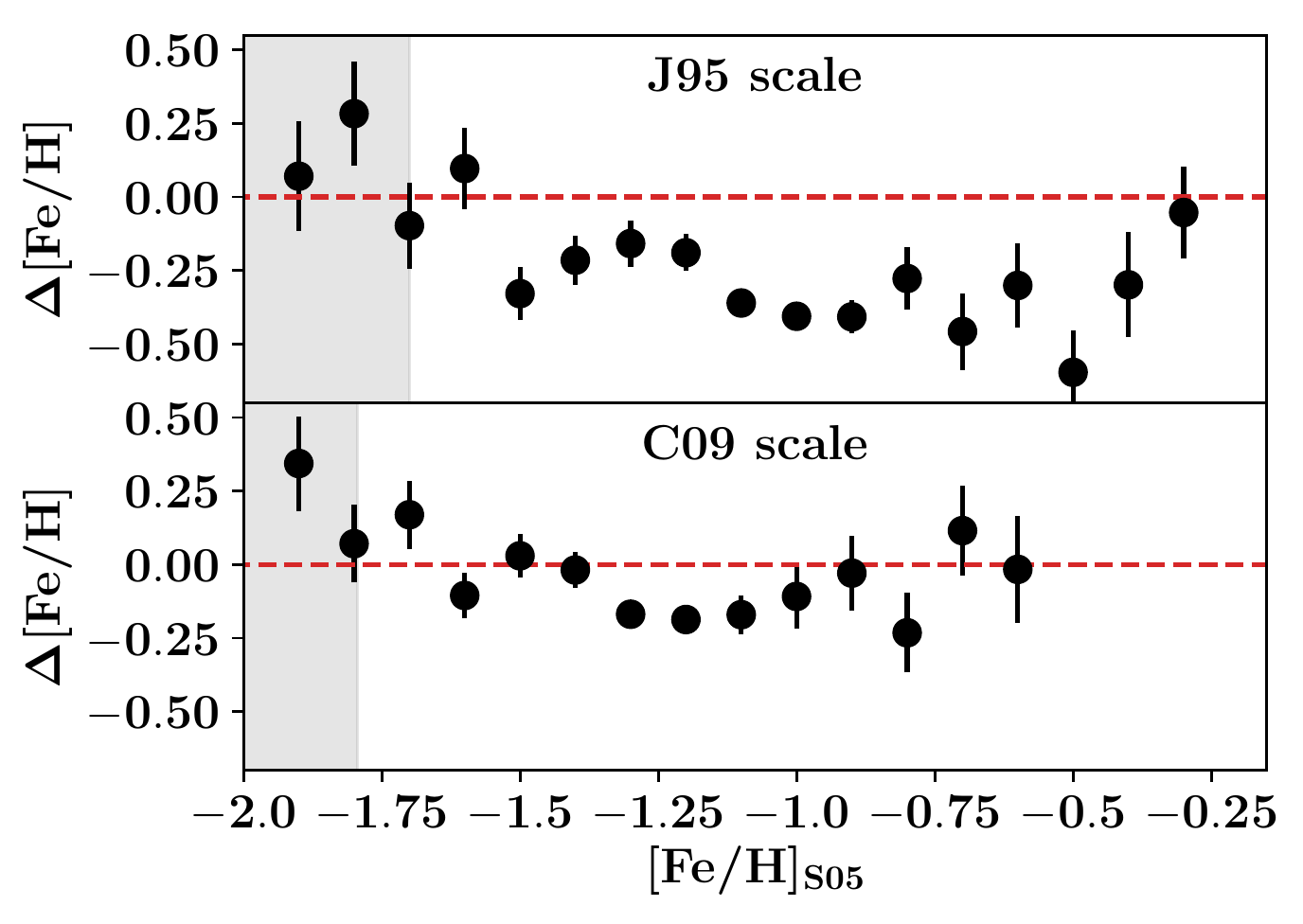}
\caption{ Median difference ($\rm [Fe/H]_{W12} - [Fe/H]_{S05}$) between the spectroscopic and photometric metallicities, as function of photometric metallicity, when the latter are anchored to the \citet[upper panel]{Jurcsik95} or the \citet[lower panel]{Carretta09b} metallicity scales. The red dashed line marks zero.The shaded grey region marks metallicities outside the photometric metallicity calibration interval.}
\label{fig:DeltaFe}
\end{figure}

The MDF obtained in this way (red histogram in Fig.~\ref{fig:Photmet}) agrees better with the spectroscopic measurement, with $[{\rm Fe/H}]_{\rm Med} = -1.24$, but does not erase the discrepancy completely. This is also reflected in a more detailed comparison between the spectroscopic and photometric metallicities. In general our uncertainties are too large to perform a meaningful comparison on a star-to-star basis. The global trend is, however, robust and we show it in Fig.~\ref{fig:DeltaFe} where we report the median difference between the W12 and S05 metallicities, as a function of the S05 metallicity. This comparison shows that the the two metallicity measurements generally follow each other. However, when S05 metallicities are left on the original J95 scale, they tend to be higher than the W12 measurements by roughly 0.3~dex. When the S05 values are transposed to the C09 scale, the agreement improves significantly. The remaining residuals are mostly below 0.15~dex. The residual structure is such that at intermediate metallicities the W12 relation gives smaller metallicities than the S05 one, while the opposite is true at the low-metallicity and high-metallicity ends of our sample. This behaviour seems compatible with the residual structure of the original W12 fit (cfr. their Fig.~4).

The W12 and the S05 calibrations have been derived independently (although a small overlap exists between the calibration samples) and are based on different observables. Therefore, the results of Fig.~\ref{fig:DeltaFe} provide a solid validation to the accuracy of both relations within a level of 0.15~dex, with the remaining difference being accounted by the choice of metallicity scale. This is particularly important for the W12 formula, that had not been verified outside the calibration sample before.

 \begin{table}
\centering
\caption{Median [Fe/H] value obtained for our RRL sample by using different prescriptions for the photometric metallicity calculation and different conversion formulas to transpose the photometric metallicities on the \citet{Carretta09b} scale.}
\begin{tabular} {llc}
\toprule
 Photometric Metallicity & Scale conversion &$[{\rm Fe/H}]_{\rm Med}$\\
 \midrule
 S05 & This work & $-1.24$  \\
 S05 & \citet{Hajdu18} & $-1.12$  \\
 S05 & \citet{Papadakis00}+C09 & $-1.40$\\
 \citet{Feast10} & C09 & $-1.25$\\
 \citet{Sarajedini06} & C09 & $-1.31$\\
 
\bottomrule
\end{tabular}
\label{tab:photmet}
\end{table}

In general, small systematic differences among these metallicity estimates should not be too surprising. Both the W12 and S05 relations have been evaluated on a sample of a few tens of objects. Metallicity scale conversions can be equally uncertain, as they rely on, often small, sample overlap among different studies. This argument is exemplified in Table~\ref{tab:photmet}, where we use several combinations of photometric metallicity prescriptions and metallicity scale conversions, from the literature, to obtain a set of photometric MDFs on the C09 scale. As shown in Table~\ref{tab:photmet}, the median metallicity obtained for our sample with these different methods can vary substantially, ranging from values just below the S05 estimate to values that very well agree with our spectroscopic determination. To capture this variability, amounting to roughly 0.35~dex, the comparison with the stellar population models will be presented both for the CaT and S05 (on the original J95 scale) metallicity values, as they are representative of the lower and upper envelope, respectively, of our metallicity confidence interval.

As a final note, the discussion above mainly concerns the accuracy of the different metallicity determinations. Regarding the metallicity precision, we can see that the photometric MDFs are noticeably narrower than the spectroscopic one, providing an upper limit to the width of the true MDF of this sample. For this reason, in the following section we will use the width of the photometric MDF as input for our stellar population models.

%----------------------------------------------------------------------------------------------------------------------------
%----------------------------------------------------------------------------------------------------------------------------
%--------------------------------------      SECTION 3     ------------------------------------------------------------
%----------------------------------------------------------------------------------------------------------------------------
%----------------------------------------------------------------------------------------------------------------------------

\begin{figure}
\centering
\includegraphics[width=0.5\textwidth]{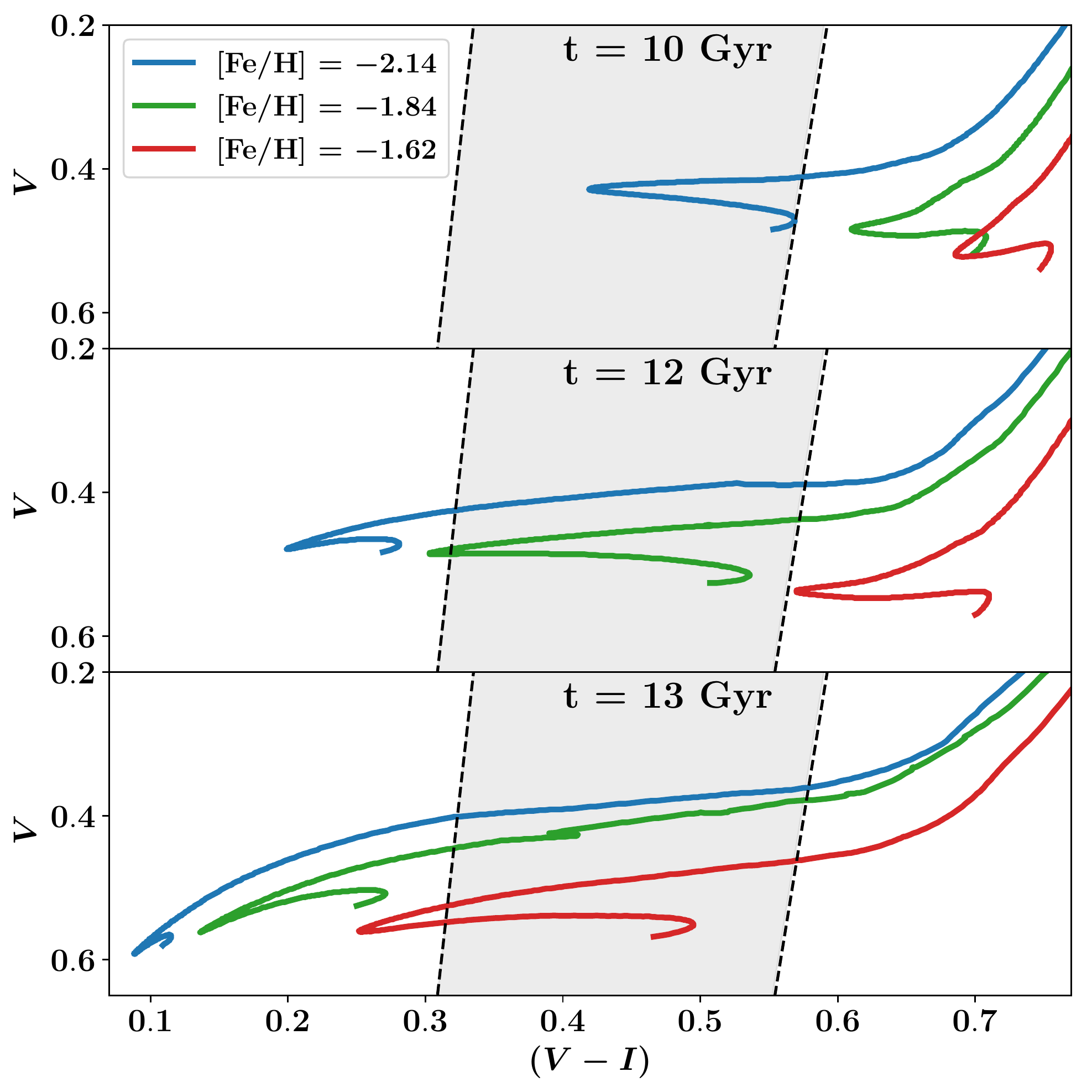}
\caption{ Schematic evolution of horizontal branch tracks in the $V$~vs~$(V-I)$ colour-magnitude diagram, as function of stellar population age. Different colours represent different metallicities. The shaded grey area marks the approximate position of the pulsation instability strip. The horizontal branch tracks are assigned to a specific age following the procedure laid out in Sect.~\ref{Model}.}
\label{fig:IS}
\end{figure}

\section{The age of RR Lyrae stars in the bulge}
\label{Model}

The main rationale behind the use of RRLs as age indicators lies in the interplay between the position of the IS in the colour-magnitude diagram and the age evolution of the HB morphology, which is schematised in Fig.~\ref{fig:IS}. As a stellar population (with low or intermediate metallicity) reaches an age greater than a few Gyr, it stops manifesting the red clump feature and develops an extended HB \citep{Girardi16}. However, not all the extent of the HB gets populated at once. At fixed metallicity, stars first ignite helium burning on the red side of the HB. As the population gets older, lower-mass stars settle on the HB at higher and higher effective temperatures. Since the IS spans a relatively narrow range of effective temperatures, HB stars of a given metallicity populate the IS only in a limited age window. Additionally, at a given age, low metallicity HB stars have bluer colours than their more metal-rich counterparts. This means that, as the stellar population ages, the IS will be populated by stars of progressively increasing metallicity. Of course, this picture is incomplete as Fig.~\ref{fig:IS} shows that, once stars of a given metallicity move to the blue of the IS, they will always cross it when they evolve towards the asymptotic giant branch. However, this transition is much more rapid than the early phases of helium-burning, so that the RRL population will typically be dominated by those stars that enter the HB inside the IS.

It should be noted that the idea that older stellar populations contain more metal-rich RRLs is not at odds with what can be expected from chemical evolution frameworks. The MDF of RRLs is in fact a biased tracer of the stellar population MDF and it can be thought as the product of the latter with an RRL production efficiency function. As the age increases, the peak of this function shifts to higher metallicities, driving the evolution of the RRL MDF.

The scenario depicted above means that we can use a stellar population synthesis approach to predict the properties of RRL populations and investigate their variations with the age and chemical abundance of their parent stellar population, as well as their robustness against the choice of theoretical ingredients used to derive them. We will therefore devote the rest of this section to build such models and compare their output with the observed properties of bulge RRLs.

%----------------------------------------------------------------------------------------------------------------------------
%--------------------------------------      MODELS      ---------------------------------------------------------------
%----------------------------------------------------------------------------------------------------------------------------

\subsection{The theoretical framework}
\label{Framework}
To calculate the HB population models we make use of the BaSTI alpha-enhanced isochrones, complemented with the BaSTI alpha-enhanced HB tracks \citep{Pietrinferni06}. This theoretical library provides a fine sampling of HB tracks (less than 0.3~dex and 0.02~$M_{\odot}$, in metallicity and mass respectively, everywhere in the covered parameter space) and it is particularly suited for synthetic HB calculations. The version of the BaSTI database we use in this work does not include the effect of atomic diffusion on the stellar structure. Although it has been shown that neglecting diffusion can lead to age overestimation of main sequence and subgiant branch stars \citep[e.g.][]{Cassisi98, Cassisi99, Dotter17}, the mixing provided by the convective envelope during the red giant branch (RGB) ascent makes the effect of diffusion much smaller for later stellar evolution phases \citep{Dotter17}. We report that the results of this paper are not significantly affected when our models are built with a stellar evolution database that includes atomic diffusion \citep{Hidalgo18}. 

The procedure to generate the synthetic stellar populations is the same adopted in \citet{Savino18a}. We refer to that paper for an in-depth description of our method and we only briefly summarise it here. For a given stellar population age and chemical composition, the corresponding isochrone, interpolated from the original BaSTI grid, is used to obtain the initial mass of stars at the tip of the RGB. We consider a range of masses slightly above that value, in proportions according to a \citet{Kroupa01} initial mass function, as our HB progenitor population. The more massive the progenitor, the earlier it leaves the tip of the RGB and the more advanced the corresponding HB star will be in its helium-burning lifetime. Stars so massive that they have completed the helium burning and evolved as white dwarfs are removed from the population. During their RGB evolution, the HB progenitors lose a significant amount of mass due to stellar winds. We take this mass loss into account to calculate the corresponding HB mass (we ignore the mass loss prescription already incorporated in the stellar isochrones and instead proceed according to Sect.~\ref{ML}). The mass, chemical composition and time since the start of the helium burning of each HB star are used to interpolate our HB track grid and calculate the corresponding structural and photometric properties. We create synthetic stellar populations with mass, at formation, of $10^7 M_{\odot}$, which results in a sample of 6000 to 8000 synthetic HB stars per model.

Once we have our synthetic HB model at hand, we decide which stars develop RRL-like pulsation. To this end, we use two different theoretical IS boundary prescriptions. The first, calculated over a large grid of structural and chemical stellar parameters, uses the linear convective pulsation code of \citet{Smolec08} and OPAL opacities \citep{Iglesias96}. The second set of models uses the IS boundaries from \citet{Marconi15}, which are based on a stellar convective  pulsation code developed by \citet{Bono94}. Regarding the blue boundary of the IS, we have to chose how to treat synthetic stars that lie in a narrow effective temperature-range called the OR zone, where both fundamental and first-overtone pulsations are stable. We are only interested in \textit{ab} pulsators but stars in this region can exist as type \textit{ab}, \textit{c} or \textit{d} RRLs and the mechanisms governing which pulsation mode manifests are currently unclear. To capture this uncertainty, we included the OR zone in the IS boundaries of \citet{Smolec08} but excluded it in those of \citet{Marconi15}, so that the effective temperature difference, at the blue boundary, between the two IS prescriptions is maximised.

These IS prescriptions determine the effective temperature boundaries for our synthetic stars to be labeled as RRLs but they pose no limitation on the brightness of potential RRLs. This is unrealistic as real RRLs are classified as such only if they show pulsation periods smaller than roughly 1 day. HB stars that cross the IS at sufficiently high luminosities have longer pulsation periods and are classified as BL Her variables \citep{Wallerstein02} instead. Therefore, we only consider as RRLs synthetic HB stars that cross the IS with $M_{I} > -0.3$. This number corresponds to the brightest absolute I band magnitude of our bulge sample, obtained through the \citet{Marconi15} period-luminosity relation, and corresponds to pulsation periods of roughly 1 day.

\begin{figure}
\centering
\includegraphics[width=0.5\textwidth]{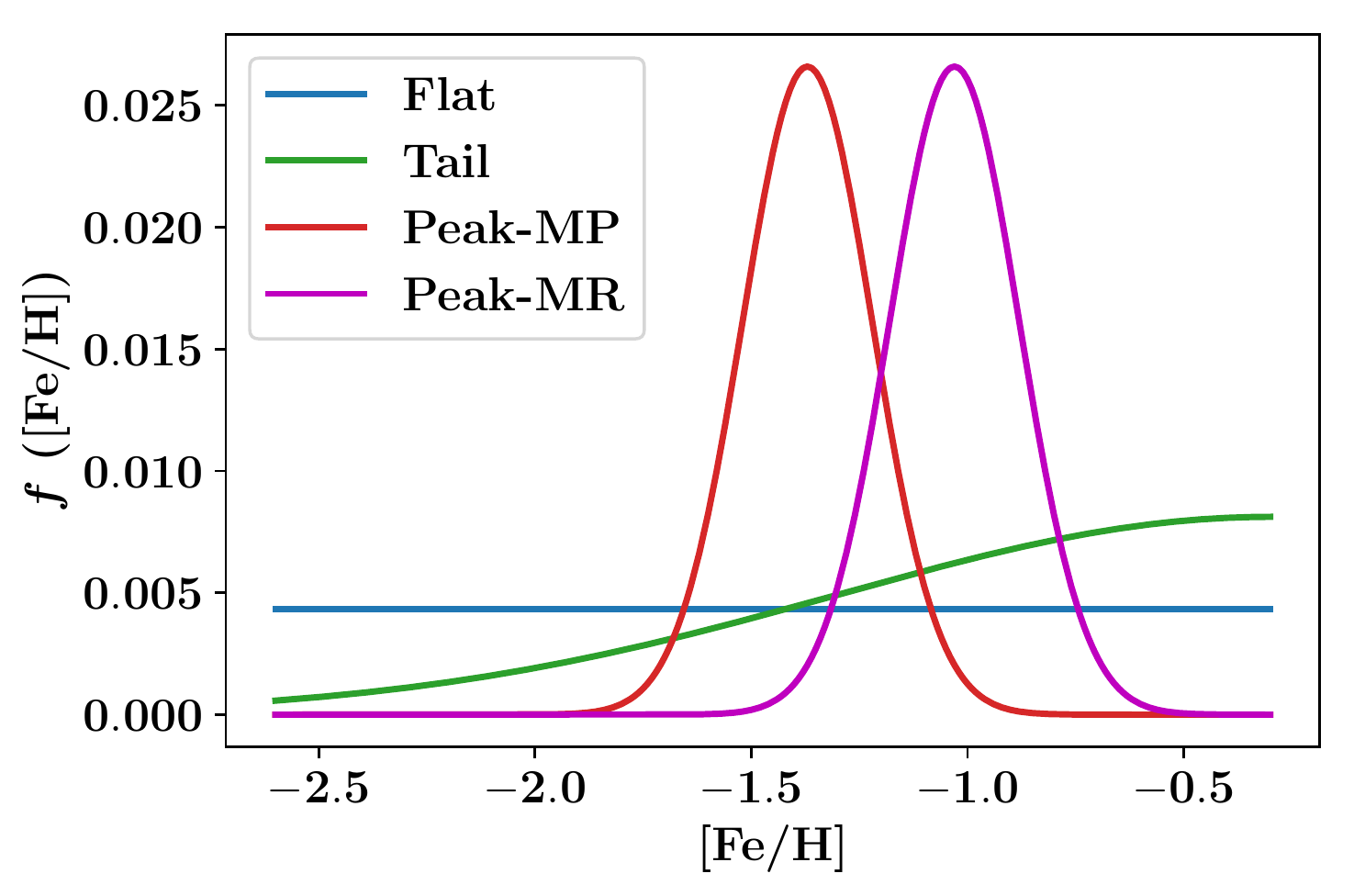}
\caption{The four metallicity distribution function models used to generate our synthetic stellar populations.}
\label{fig:MDFs}
\end{figure}

%----------------------------------------------------------------------------------------------------------------------------
%--------------------------------------      STELPOPS      ------------------------------------------------------------
%----------------------------------------------------------------------------------------------------------------------------

\subsection{The stellar population parameters}
The next step is to specify the distributions of age and chemical composition that will be used to construct our stellar population models. Stars in the bulge have been shown to be alpha-enhanced up to [Fe/H] = $-0.4$ \citep[e.g.][]{Hill11,Bensby17,McWilliam16,Rojas-Arriagada17}, so we assume $[\alpha/{\rm Fe}] = 0.4$ in our calculations. We also assume that the helium abundance of our populations follows the canonical scaling with metallicity included in theoretical stellar models \citep{Basti}. For our models this ranges from Y~=~0.245, at [Fe/H]~=~$-2.5$, to Y~=~0.273, at [Fe/H]~=~$-0.3$. In Sect.~\ref{helium} we will explore the effect of departing from this assumption. We assume that our stellar populations formed in a relatively short burst of constant star formation, lasting 500~Myr. We create different models for mean ages  that range from 9.5~Gyr to 14.0~Gyr, with a step of 0.5~Gyr.

We also need to adopt a MDF for our stellar population. The MDF of our RRL sample, in fact, is biased by the RRL production efficiency and it is not representative of the true MDF of the underlying stellar population. Although great progress has been made in measuring the bulge MDF \citep[e.g.][]{Ness13,Rojas-Arriagada17,Zoccali17}, the precise shape of the metal-poor tail is still poorly constrained \citep[e.g.][]{Koch16,Arentsen20}, due to its minor contribution to the total bulge mass. We calculated our models using 4 different MDF assumptions, shown in Fig.~\ref{fig:MDFs}. These continuous MDFs are intended to represent a range of possible scenarios that, while not necessarily reflecting the physics of bulge formation, are different enough to capture the dependency of our results on the MDF uncertainty. The first model is a flat distribution over our entire metallicity range (blue curve, referred from now on as the Flat model) and it reflects a uniformed prior on the metallicity distribution of the RRL stellar population. The second model is the metal-poor tail of a Gaussian distribution that peaks at [Fe/H] = $-0.3$, with standard deviation of 1~dex (green curve, Tail model), compatible with the idea that the RRL stellar population is the metal-poor extrapolation of the main, more metal-rich, stellar population of the bulge. The final two models assume that the underlying MDF matches exactly the observed RRL MDF and are built with a Gaussian profile, with standard deviation of 0.15~dex (derived from the photometric metallicity distribution) and peaking at the median [Fe/H] of our CaT (red curve, Peak-MP) and S05 (magenta curve, Peak-MR) MDFs, respectively.

\begin{figure}
        \subfloat[][]
	{\includegraphics[width=0.5\textwidth]{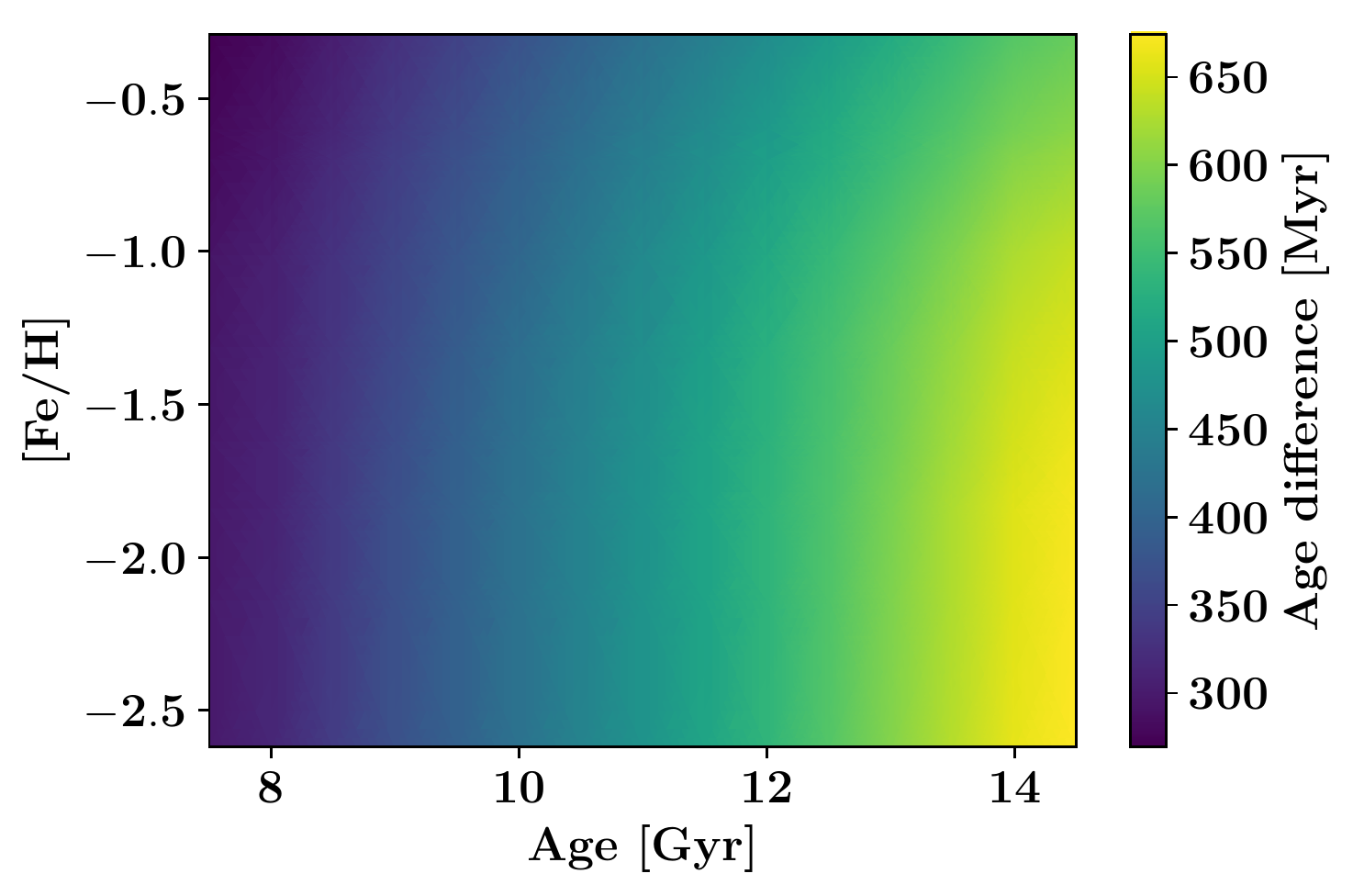} \label{fig:shift}} \quad
	 \subfloat[][]
	{\includegraphics[width=0.5\textwidth]{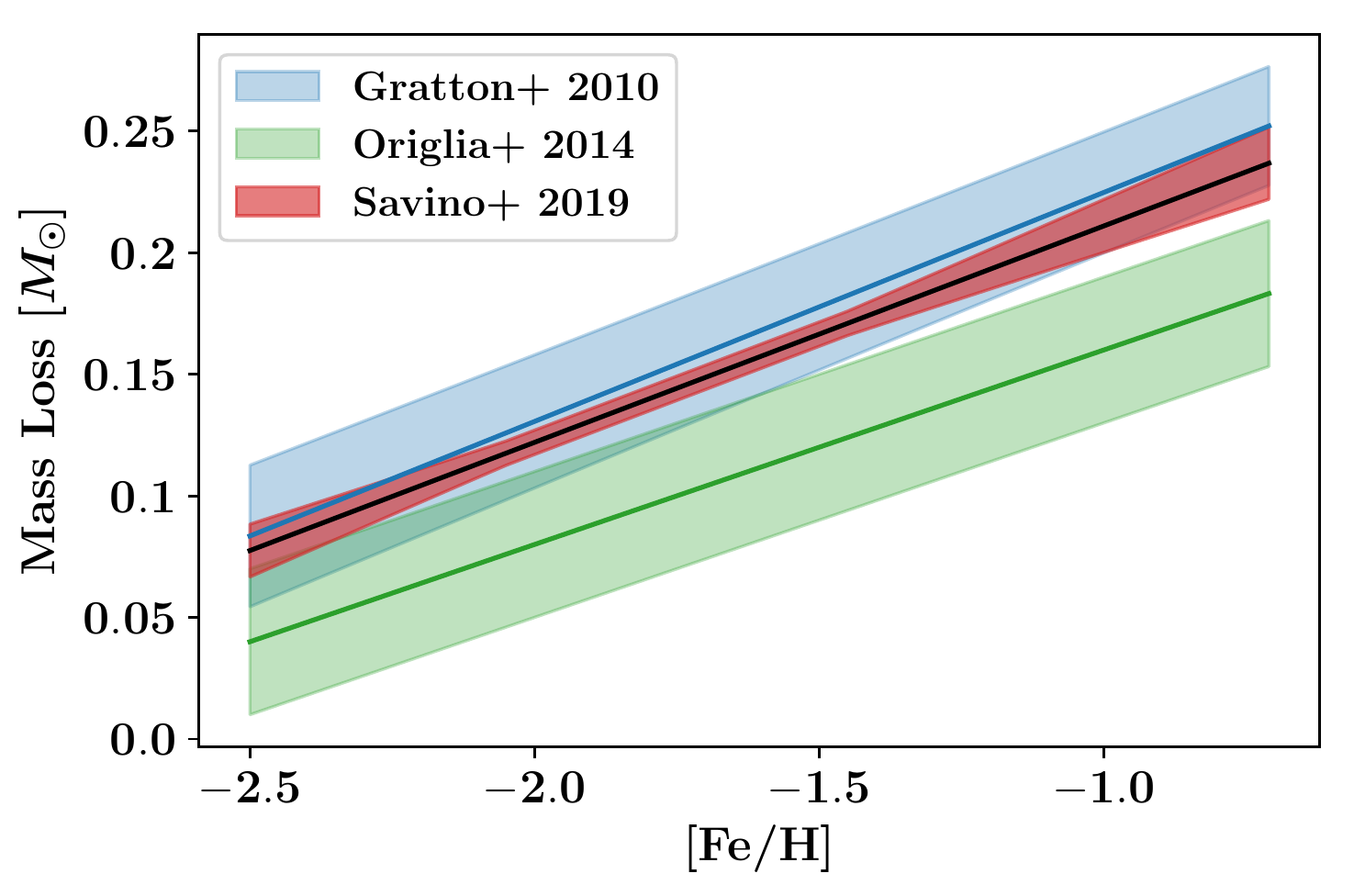} \label{fig:ML}} \quad
  \caption{a) Systematic error, as a function of stellar population age and metallicity, introduced in horizontal branch based ages by a systematic error of $0.01 M_{\odot}$ in the red giant branch mass loss. b) Red giant branch mass loss relation of \citet[black solid line and red shaded band]{Savino19a}, adopted in this paper, alongside the relations of \citet[blue]{Gratton10} and \citet[green]{Origlia14}. The band width represents mass loss uncertainties.}
    \label{fig:rad}
\end{figure}

%----------------------------------------------------------------------------------------------------------------------------
%--------------------------------------      RGB ML      -----------------------------------------------------------------
%----------------------------------------------------------------------------------------------------------------------------

\subsection{The red giant branch mass loss}
\label{ML}
The final ingredient needed for the generation of synthetic HB models is a prescription for the mass loss along the RGB evolution. The total amount of mass that a star loses before entering the helium-burning phase is a necessary quantity to provide a mapping between the mass of a given HB star and the initial mass of the corresponding stellar model at the tip of the RGB that, at fixed chemical composition, is related to the age of the stellar population.

Because old stellar populations evolve very slowly, a small difference in the estimated RGB-tip mass results in a large difference in the inferred age. This means that we need to know the RGB mass loss with great precision in order to limit our age uncertainties. This is illustrated in Fig.~\ref{fig:shift} that shows, for different age and metallicity combinations, how much the age inferred from the HB properties would change by changing the amount of RGB total mass loss by 0.01~$M_{\odot}$. This figure shows that even an uncertainty of a few hundredth of solar mass would result in age uncertainties well above 1~Gyr. This strong degeneracy between RGB mass loss and stellar population age has been the main factor hindering the use of helium-burning stars to achieve precise stellar population dating and is, for instance, the reason why \citet{Lee92} could not place bulge RRLs on an absolute age scale, resorting instead to a comparative study with the halo RRL population. The RGB mass loss has been, indeed, for long time a particularly difficult parameter to measure.

\begin{figure}
\centering
\includegraphics[width=0.5\textwidth]{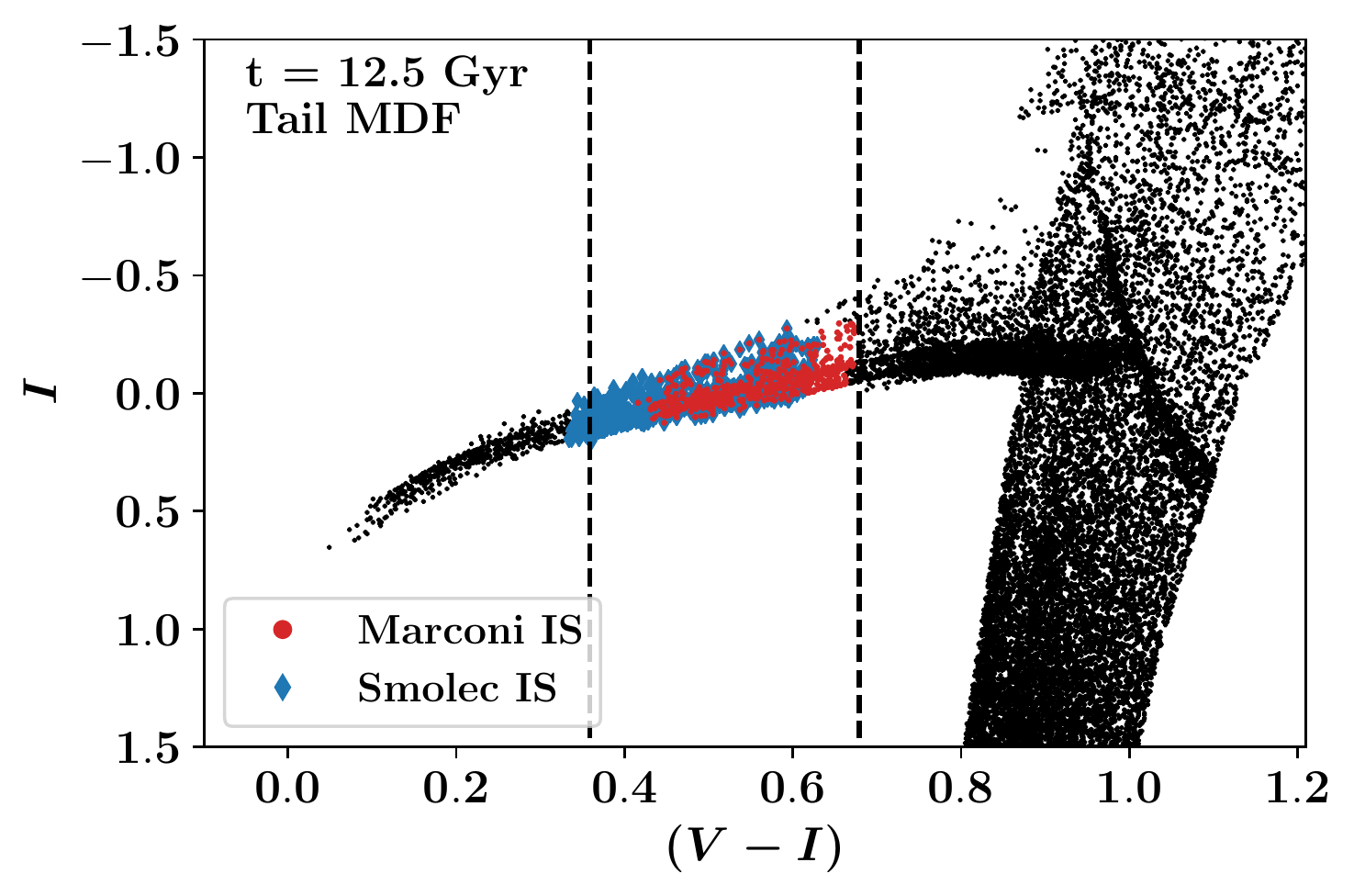}
\caption{Representative $I$~vs~$(V-I)$ colour magnitude diagram for our synthetic population models. This population has been generated using the Tail metallicity distribution function model and a mean age of 12.5~Gyr. Red circles and blue diamonds represent stars flagged as RR Lyrae using the \citet{Marconi15} and \citet{Smolec08} instability strip models, respectively. The dashed black lines mark the intrinsic colour range of our bulge RR Lyrae sample.}
\label{fig:HB}
\end{figure}

A common choice in stellar population models is to use the relation developed by \citet{Reimers75}, to link the physical properties of a star to its mass loss rate. Not only this relation was obtained from a limited dataset of stars in a mass and metallicity regime very different from our sample, it also requires the calibration of a free efficiency parameter, $\eta$. The values of $\eta$ typically found in literature range from 0.2 to 0.5 \citep[e.g.][]{Miglio12,Lei13,McDonald15}, therefore changing significantly the total RGB mass loss. Empirical estimates of the total mass loss of RGB stars in Galactic globular clusters have been provided by \citet{Gratton10} and \citet{Origlia14}, but have uncertainties of several hundredths of solar mass. Recently a very precise measurement has been obtained in the Tucana dwarf spheroidal galaxy by \citet[][hereafter S19]{Savino19a}, with uncertainties below 0.01~$M_{\odot}$ over a large range of metallicity. This level of precision is allowed by the use of a novel self-consistent modelling technique on the entire colour-magnitude diagram \citep{Savino18a} and by the smaller number of parameters affecting HB stars in dwarf galaxies compared to globular clusters \citep{Bastian18}. This S19 relation, shown in red Fig.~\ref{fig:ML} is our choice to compute the stellar population models. We adopt a mass loss dispersion at fixed metallicity of 0.005~$M_{\odot}$, as large values are disfavoured by the HB properties of globular clusters and dwarf spheroidal galaxies \citep[][S19]{Caloi08}.

\begin{figure*}
        \subfloat[][]
	{\includegraphics[width=0.5\textwidth]{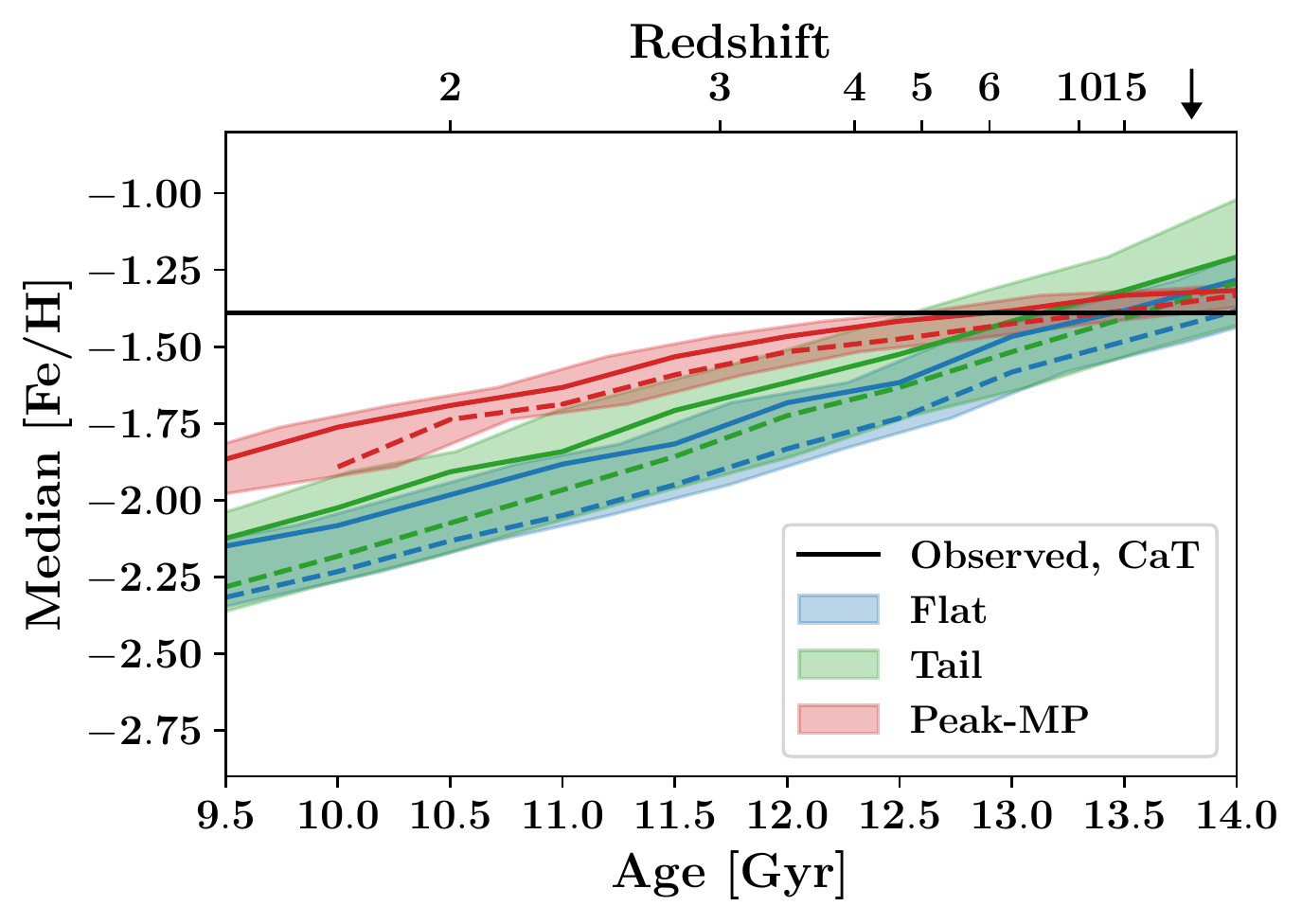} \label{fig:modellow}} \quad
	 \subfloat[][]
	{\includegraphics[width=0.5\textwidth]{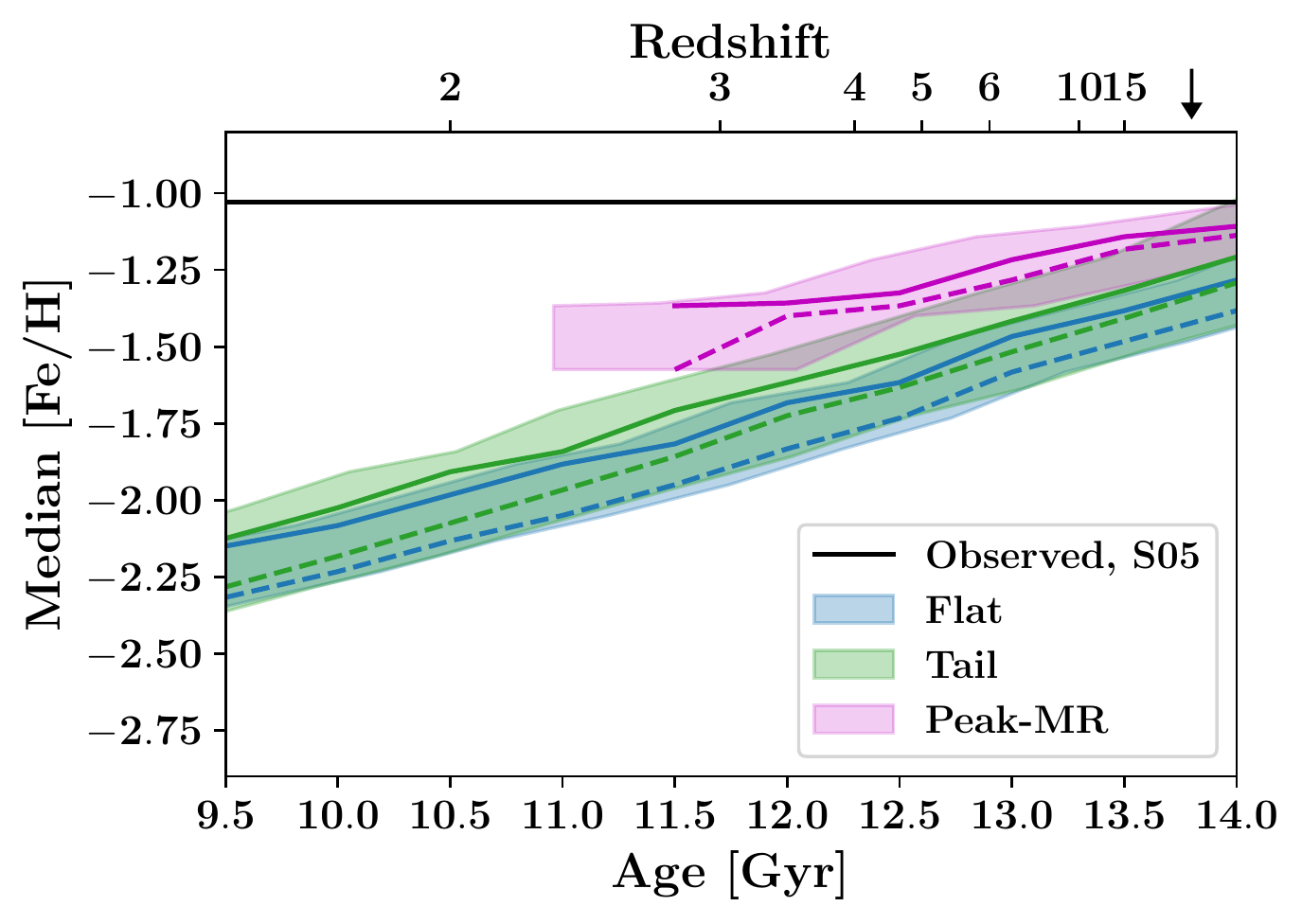} \label{fig:modelhi}} \quad
  \caption{Median [Fe/H] abundance, as a function of age, for our RR Lyrae population models. Different colours refer to different metallicity distribution functions. Solid lines refer to models with the \citet{Marconi15} instability strip prescription and dashed lines refer to models with the \citet{Smolec08} instability strip prescription. The shaded area indicates the uncertainties related to the red giant branch mass loss. The solid black line shows the median metallicity measured for our sample of bulge RR Lyrae stars. The arrow marks the age of the Universe, according to the cosmological parameters of \citet{Planck18}. a) Comparison with the spectroscopic metallicities. b) Comparison with the \citet{Smolec05} photometric metallicities.
  }
    \label{fig:Metage}
\end{figure*}

While the S19 prescription is currently the most appropriate option to model the mass distribution of helium-burning stars in the bulge, there are two caveats that are worth discussing. The first one is that, since the nominal uncertainties on the S19 relation are so small, even small systematic effects can become the dominant source of uncertainty. These can stem from, e.g., imperfections in the stellar evolution models or limitations in the approach used to derive the relation. Quantifying these types of systematics is often a very difficult endeavour. An additional reason why we adopted the S19 relation is that it has been measured using exactly the same set of stellar tracks and the same stellar population modelling that are used in this paper, ensuring consistency that should minimise at least some of the above mentioned effects.

The second point is that the S19 relation has been derived in a completely different type of environment compared with the bulge, viz. a dwarf spheroidal galaxy. One could therefore question whether the same relation should apply to the stellar population we are considering here. This is a legitimate concern and it is difficult to verify, as the studies on the RGB mass loss still lack the sample size and diversity necessary to characterise the environmental variability of this quantity. Given these premises, we have to resort to the reasonable assumption that, at fixed mass and chemical composition, the evolution of a star in isolation should not be dependent on the specific environment in which that star formed, as long as that environment is not so dense that star-to-star interactions start playing a role (which is neither the case for dwarf spheroidal galaxies or for the Galactic bulge).

%----------------------------------------------------------------------------------------------------------------------------
%--------------------------------------      AGE     -----------------------------------------------------------------------
%----------------------------------------------------------------------------------------------------------------------------

\subsection{Age-metallicity relation of RR Lyrae populations}
With the procedure described above, we can calculate synthetic HBs, and the associated RRL populations, very efficiently for a range of physical and theoretical inputs. An example of our HB models is shown in Fig.~\ref{fig:HB}, where we highlighted the synthetic RRLs obtained with the two IS prescriptions. The mean colour range enclosing 95\% of our bulge RRL sample is also reported. This intrinsic colour is calculated from the $M_V$ and $M_I$ absolute magnitude relations provided in \citet{Catelan04}, which are in turns dependent on the observed period and photometric metallicities. Our synthetic RRLs span similar colours to the observed RRLs. Although neither IS prescriptions reproduces perfectly the observed colour range, this is not surprising, as the comparison critically depends on the treatment of the OR zone and on the uncertain metallicity values of the observed RRLs.

From these models, we can extract the metallicity of the synthetic RRLs and quantify how it depends on the stellar population parameters and the other model ingredients. In principle, we could extract the complete MDF of our model RRLs. However, the detailed comparison between the model and observed MDFs would be very sensitive to, e.g., incorrect characterisation of the measurement uncertainties, selection biases in the data or imperfections in our theoretical framework. We instead prefer to trace only the median metallicity of our synthetic population, as this is likely to be more robust against the effects mentioned above and, as it will be shown, already provides us with a powerful tool for RRL dating.

The median RRL metallicity for different assumed MDFs and IS prescriptions is shown, as function of stellar population age, in Fig.~\ref{fig:Metage}, where the shaded regions represent the standard deviation of every model, computed with 300 different realisations of the RGB mass loss, according to the uncertainties quoted in S19. As we anticipated, there is a very tight relation between the age of a given stellar population and the typical metallicity of its RRL population, with the metallicity becoming higher for older stellar populations. The uncertainties on the RGB mass loss and the IS boundaries introduce a similar amount of scatter in the models.

The choice of MDF has a more moderate impact, with the exception of the two Peak models at young ages, where they tend to manifest higher metallicities compared to the Flat and Tail models. This can be explained as following. At old ages the RRL production efficiency peaks at metallicities that are very well populated in all models. This results in very similar metallicity distributions for the variable stars. At young ages, the production efficiency drops dramatically for the Peak models, as only the few stars in the metal-poor tail of the population can become RRLs. Since the Peak MDFs drop steeply away from the maximum, this skews the RRL metallicity distribution towards more metal-rich values, compared to models that have more evenly distributed metallicities. An opposite effect would have occurred had we assumed a narrow MDF peaking at very metal-poor values. In that case, the median RRL metallicity would have agreed with the Flat and Tail models at young ages and it would have risen more slowly at old ages, when the RRL production peaks in the scarcely populated metal-rich tail.

\begin{figure*}
\centering
\includegraphics[width=0.9\textwidth]{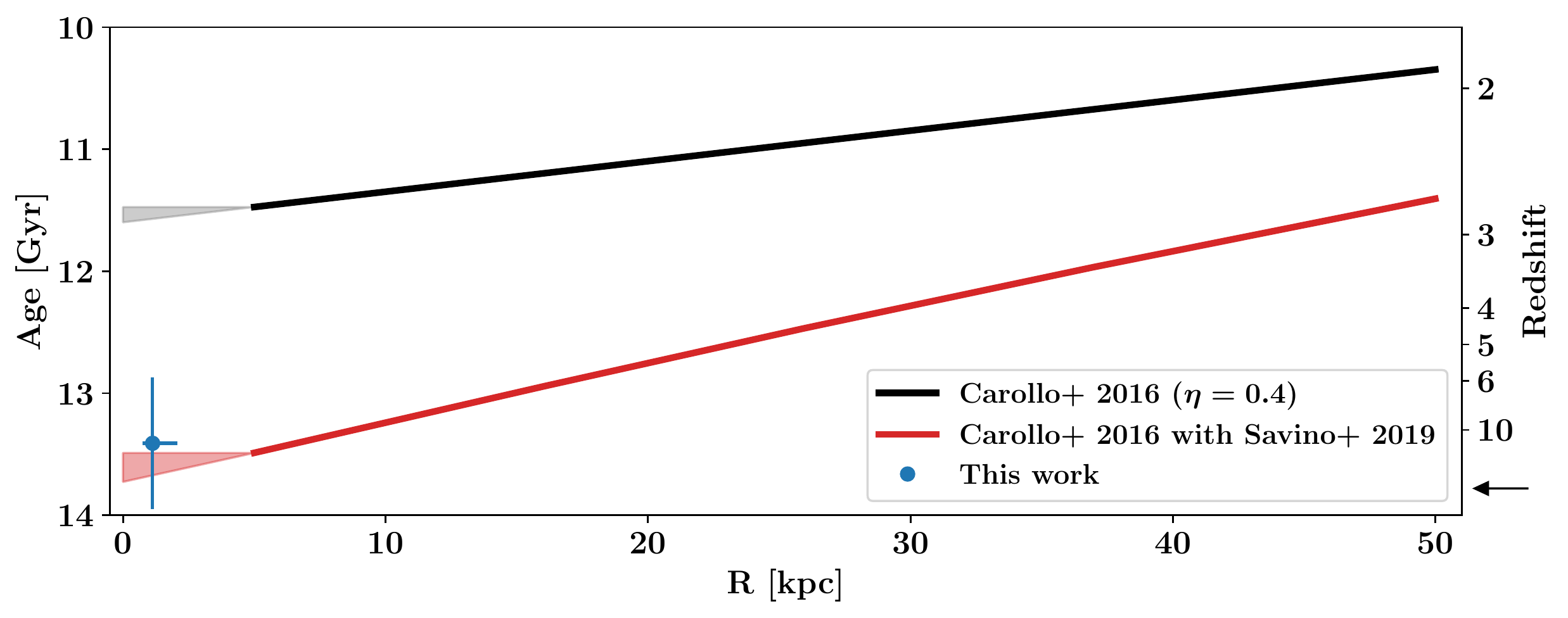}
\caption{Radial age profile derived by \citep{Carollo16} for the blue horizontal branch population of the stellar halo (black solid line). The red line shows the same profile once corrected for the red giant branch mass loss used in this paper. The shaded areas show the profile extrapolations assuming either a flattening or a steady gradient. The blue dot marks the age obtained for the bulge RR Lyrae population. The arrow marks the age of the Universe, according to the cosmological parameters of \citet{Planck18}.}
\label{fig:gradient}
\end{figure*}

The coherent behaviours of these models under the range of assumptions enable us to derive a generalised relation between the median metallicity of RRLs and the parent stellar population age. This has been obtained by fitting our model ensamble with a second order polynomial, in the $-2.3<[{\rm Fe/H}]<-1.3$ range, resulting in:
\begin{equation}
t=26.62+13.12\cdot[{\rm Fe/H}]+2.50\cdot[{\rm Fe/H}]^2 \quad [{\rm Gyr}],
\label{eq:absage}
\end{equation}
\begin{equation}
\sigma_t = -3.19-4.69\cdot[{\rm Fe/H}]-1.38\cdot[{\rm Fe/H}]^2 \quad [{\rm Gyr}],
\end{equation}
where the uncertainty has been obtained from the scatter of our models and ranges between 300 and 800~Myr, depending on metallicity. %We note that the models differ mainly through a zero point offset but have very similar slopes. We also note that, while the uncertainties in parameters such as the RGB mass loss and the boundaries of the IS introduce uncertainties in the absolute age scale we derive, the physics of this processes applies in the same way to different stellar populations. We can therefore derive a much more precise differential relation, to estimate the age difference of RRL populations with differences in their median metallicity:
%\begin{equation}
%\frac{dt}{d[Fe/H]} = 45.005+40.922\cdot[Fe/H]+10.357\cdot[Fe/H]^2,
%\label{eq:diffage}
%\end{equation}
%\begin{equation}
%\sigma_{\frac{dt}{d[Fe/H]}} = 27.178+25.997\cdot[Fe/H]+6.287\cdot[Fe/H]^2.
%\label{eq:diffunc}
%\end{equation}
%This differential relation can be applied in the range $-2.3<[Fe/H]<-1.6$.

We stress that this prescription should be used with caution, as it relies on the assumptions we made in constructing the stellar population models. It is valid only in the quoted metallicity range and under the assumption that the underlying MDF of the stellar population is not drastically different from the ones adopted here. Finally, we stress that this relation is meant to be used only with the median metallicity of a population of RRLs and should never be used to derive the age of a single RRL star.

\subsection{The age of the inner spheroid}
From these stellar population models we can see that RRLs with $[{\rm Fe/H}] \gtrsim -1.5$ are only compatible with very old stellar populations. In particular, we can use the results from Sect.~\ref{Obs} to infer the age of the stellar spheroid in the bulge. We first look at the metallicities obtained from the CaT spectroscopy, with a median [Fe/H] value of $-1.39$. From the models reported in Fig.~\ref{fig:modellow}, we obtain an age of $13.41\pm0.54$~Gyr, corresponding to a formation redshift of $z=11.6\pm_{5.6}^{-}$ \citep[adopting][ cosmological parameters]{Planck18}. We obtain this figure by direct interpolation of our stellar population models and not through Eq.~\ref{eq:absage}. This extremely ancient age is comparable to the age of the oldest known stars in the Milky Way \citep[e.g.][]{Frebel07,Vandenberg14} and suggests that RRLs in the bulge were among the first stars to form in what is now the Milky Way galaxy. 

We also see from Fig.~\ref{fig:modelhi}, that adopting more metal-rich determinations for the bulge RRLs introduces tension in the comparison with our models. Reproducing the most metal rich value of $[{\rm Fe/H}]_{\rm Med} = -1.03$, obtained through the S05 photometric metallicities, would require most of our models to have ages significantly older than the age of the Universe and only the Peak-MR stellar population models are marginally compatible with this measurement. We will return on this issue in Sect.~\ref{Alternatives}, were we will explore alternative scenarios for the formation of metal-rich RRLs. For now we conclude that, within the modelling framework presented here, bulge RRL metallicity values in the upper region of our confidence interval further strengthen the inference of an extremely old age for this stellar population, placing it among the firsts to have formed in the Milky Way.

%----------------------------------------------------------------------------------------------------------------------------
%--------------------------------------      OUTER HALO      --------------------------------------------------------
%----------------------------------------------------------------------------------------------------------------------------

\subsection{Comparison with the outer spheroid}
Having provided the first quantitative determination for the stellar age of the innermost stellar spheroid of the Milky Way, the question naturally arises of how this measurement compares with what is known about the large scale stellar spheroid of the Galaxy, i.e. the stellar halo\footnote{Although the stellar halo itself has been shown to consist of an inner and outer component \citep[e.g.][]{Hartwick87,Carney96,Carollo07}, we will here refer to the entirety of the stellar halo outside the bulge as the ``outer spheroid".}. Evidence that a radial colour gradient exists in the halo population of blue HB (BHB) stars and its interpretation as a mean age variation was reported already by \citet{Preston91}. Both the sample and the methodology were incrementally improved by \citet{Santucci15}, \citet{Das16} and \citet{Carollo16}, with the latter study providing an age gradient for the halo BHB population that covers a range of Galactocentric distances from 5 to 50~kpc. Such gradient is shown as a black line in Fig.~\ref{fig:gradient}. Conversely, our RRL sample has a median Galactocentric distance of 1.13~kpc, spans a range of 0.75 to 2.07~kpc from the 15.87th to the 84.13th percentile, and it is represented as the blue point.

A direct comparison between the two results would point at strong differences between the bulge stellar spheroid and the halo, which appears to be younger by approximately 2 Gyr. However, the reason for this difference lies in the procedure adopted to derive the stellar ages. Being based on the modelling of HB stars, the age scale derived in \citet{Carollo16} critically depends on the choice of an RGB mass loss prescription, similarly to our analysis. In particular, the authors adopt a Reimers mass loss law, with an efficiency parameter $\eta=0.4$. This high efficiency parameter results in mass loss amounts that are relatively similar to the ones prescribed by S19 at high metallicities. At metallicities typical of the stellar halo, however, the use of $\eta = 0.4$, results in substantially higher values compared to S19. This makes the inferred ages significantly younger.

To provide a consistent comparison with our results, we must transform the halo measurements to the age scale used in this work. We used the map provided in Fig.~\ref{fig:shift} and the BaSTI stellar isochrones to infer an age dependent correction factor and account for the RGB mass loss difference between a Reimers law with $\eta=0.4$ and the S19 law. We calculate this factor at [Fe/H]~$=-1.75$, which is the typical metallicity of the halo BHB sample. We apply this factor to the gradient of \citet{Carollo16} and obtain a new profile, shown in Fig.~\ref{fig:gradient} as a red line.

Adopting the S19 mass loss, the halo age profile gets significantly older and steeper, so that in the inner 10~kpc the halo stellar ages are compatible with those derived here for the bulge stellar spheroid. With the caveat of the sizeable error bars, there seems to be a reasonable continuity between the stellar ages of the inner and the outer spheroids, provided the radial profile flattens inside 5~kpc. This flattening is almost necessarily bound to exist, as the stellar age inferred at 5~kpc is already the oldest meaningful age for the formation of population II stars. This is supported by the spatial properties of our bulge RRL sample. Using distances from \citet{Kunder20}, in fact, we report the absence of any statistically significant radial gradient, in either the spectroscopic or photometric metallicities, between 0.5 and 4~kpc from the Galactic centre, which also rules out a significant age gradient. We note that a metallicity gradient in bulge RRLs was reported by \citet{Kunder08} and \citet{Pietrukowicz15}. However, the reported gradients are very mild, with the latter study quantifying it at 0.02~dex/kpc. The nearly constant metallicity profile seems to break at around 5-6~kpc, where the mean RRL metallicity starts decreasing more steeply \citep{Suntzeff91,Pietrukowicz16}.

This continuity in age and metallicity is not the only connection between the inner and the outer stellar spheroids. Using stellar counts, \citet{Perez-villegas17} showed that the radial density profiles of bulge and halo RRLs are compatible with each other, both in slope and normalisation. This degree of spatial, chemical and chronological coherence between the two structures gives support to the hypothesis that the spheroidal population in the bulge is in fact the extrapolation at small radii of the Galactic halo, as first suggested by \citet{Minniti99}. The existence of a breaking point in the halo properties at around 5~kpc, if confirmed, would however open to the possibility that the innermost stellar halo is somehow different from its large-scale counterpart and would raise questions on what are the mechanisms governing this ulterior transition.

\begin{figure}
\centering
\includegraphics[width=0.5\textwidth]{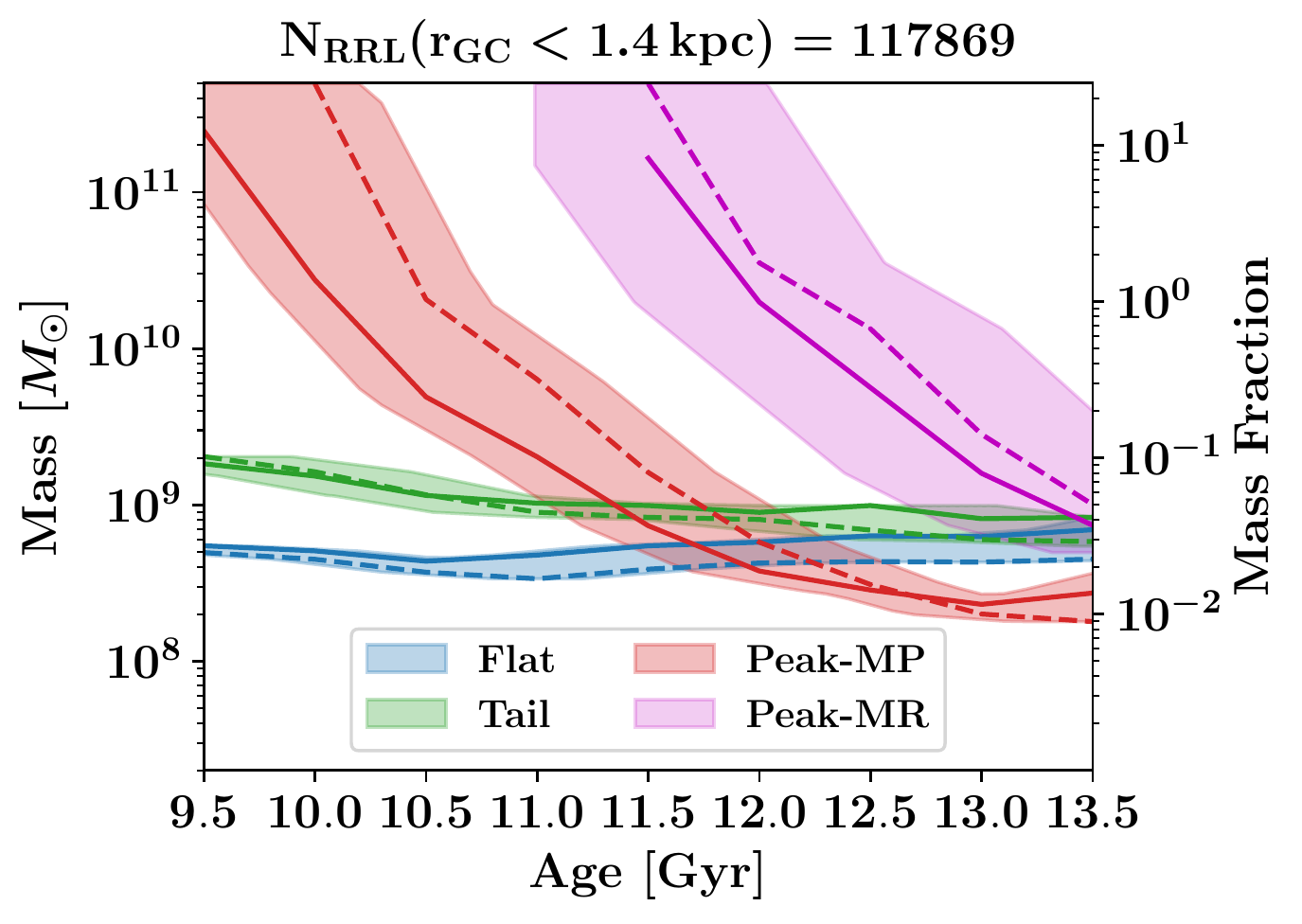}
\caption{Mass of the Milky Way central spheroidal population, obtained from the RR Lyrae production efficiency of our models and the total number of bulge RR Lyrae estimated in Sect.~\ref{Mass}. The colour and symbol coding of our models is analogous to Fig.~\ref{fig:Metage}. The right y axis shows the bulge stellar mass fraction contained in the old spheroid, assuming the bulge total mass from \citet{Valenti16}.}
\label{fig:mass}
\end{figure}
%----------------------------------------------------------------------------------------------------------------------------
%--------------------------------------      MASS     --------------------------------------------------------------------
%----------------------------------------------------------------------------------------------------------------------------

\subsection{The mass of the inner spheroid}
\label{Mass}
A natural outcome of our stellar population models is the ability to infer the total stellar mass contained in the central spheroid. For each of our models we can calculate the integrated RRL production efficiency, i.e. the number of RRLs produced per unit stellar mass, and multiply it by the estimated total number of RRLs in the bulge to obtain the mass of the underlying stellar population. We estimate the number of RRLs from the density profile of \citet{Perez-villegas17}, integrated within a sphere of 1.4~kpc of radius. This corresponds to a radius of roughly 10~deg in the sky, which is what is typically defined as the bulge region.

As the slope of the power law density profile is very close to $-3$, the total number of RRLs is influenced significantly by the shape of the profile at small radii. While the density has been measured to be an unbroken power law down to 200 pc from the Galactic centre, we do not have data for smaller radii. To avoid the divergence of the integral, we assume that the stellar density flattens at a $r_{\rm core} = 100$~pc, yielding a total number of RRLs of roughly 118,000. Letting $r_{\rm core}$ vary between 200~pc and 5~pc \citep[roughly the radius of the nuclear star cluster,][]{Schodel14}, changes the number of RRLs in the bulge by about a factor of 2. 

The total mass of the RRL parent population is shown for our models in Fig.~\ref{fig:mass}. Models with a broad MDF correspond to a stellar mass between a few $10^8$ and $10^9 M_{\odot}$, with almost no dependance on the stellar population age. These values correspond to a 2-5\% mass fraction, assuming a bulge total stellar mass of $2\cdot10^{10} M_{\odot}$ from \citet{Valenti16}. This value is compatible, albeit a little higher, with the 1\% estimate provided by \citet{Kunder16}. Our slightly higher mass fraction, however, refers to the entirety of the stellar population included in our models, including those stars that evolve on the blue and red HB, while the estimate of \citet{Kunder16} refers only to the subpopulation that forms RRLs. The Peak models, instead, show a dramatic age dependence. This is due to the drastic decrease of RRL production efficiency at young ages, so that much more stellar mass is required to produce the same number of variable stars. The existence of this massive stellar population is clearly at odds with most of the chemodynamical data available for the bulge. Therefore this proves that, in order for the spheroid to have a narrow MDF, it must have an age where RRL production is near the maximum of efficiency for its metallicity.

%----------------------------------------------------------------------------------------------------------------------------
%----------------------------------------------------------------------------------------------------------------------------
%--------------------------------------      SECTION 4      ------------------------------------------------------------
%----------------------------------------------------------------------------------------------------------------------------
%----------------------------------------------------------------------------------------------------------------------------

\section{Formation channels for metal-rich RR Lyrae stars}
\label{Alternatives}

As we saw in the last section, the stellar population synthesis framework we built suggests that the metallicity of the RRL population in the bulge is close to the highest that can be obtained in a Hubble time. In fact, adopting values for the median metallicity of our sample that are near the upper end of our confidence interval ($[{\rm Fe/H}]_{\rm Med} \simeq -1.0$), introduces tension with the metallicity range allowed by the models of Sect.~\ref{Model}. While this could be interpreted as an indication that the RRL sample is slightly more metal-poor than measured through the S05 method, an independent confirmation (possibly based on high-resolution spectroscopy) that the peak of the distribution lies indeed around $[{\rm Fe/H}] = -1.0$, would establish that the source of this tension lies in our stellar population models.

Furthermore, we are still required to justify the existence of RRLs with $[{\rm Fe/H}] \geq -1.0$ to explain the full extent of the MDF shown in Fig.~\ref{fig:MDF}. While the existence of a tail of metal-poor RRLs can easily be explained, either by introducing a younger subpopulation or by having a population of coeval BHBs evolve toward the asymptotic giant branch, stars close to solar metallicity are not expected to cross the IS at any point during their helium-burning life, at least under the assumptions we adopted so far in this paper. Yet, metal-rich RRLs are not only present in the Galactic bulge \citep[e.g.][]{Hansen16} but are also known to exist in the disc of the Galaxy \citep[e.g.][]{Layden95a,Liu13,Zinn20,Prudil20} and their formation mechanisms have long been source of puzzling \citep[e.g.][]{Taam76,Layden95b,Marsakov19}. In this section we will explore some of the mechanisms that could explain the existence of metal-rich RRLs and how these scenarios fit into what is known about the central RRL population of the Galaxy.

%----------------------------------------------------------------------------------------------------------------------------
%--------------------------------------      MASS LOSS      -----------------------------------------------------------
%----------------------------------------------------------------------------------------------------------------------------

\subsection{Enhanced RGB mass loss}
\begin{figure}
\centering
\includegraphics[width=0.5\textwidth]{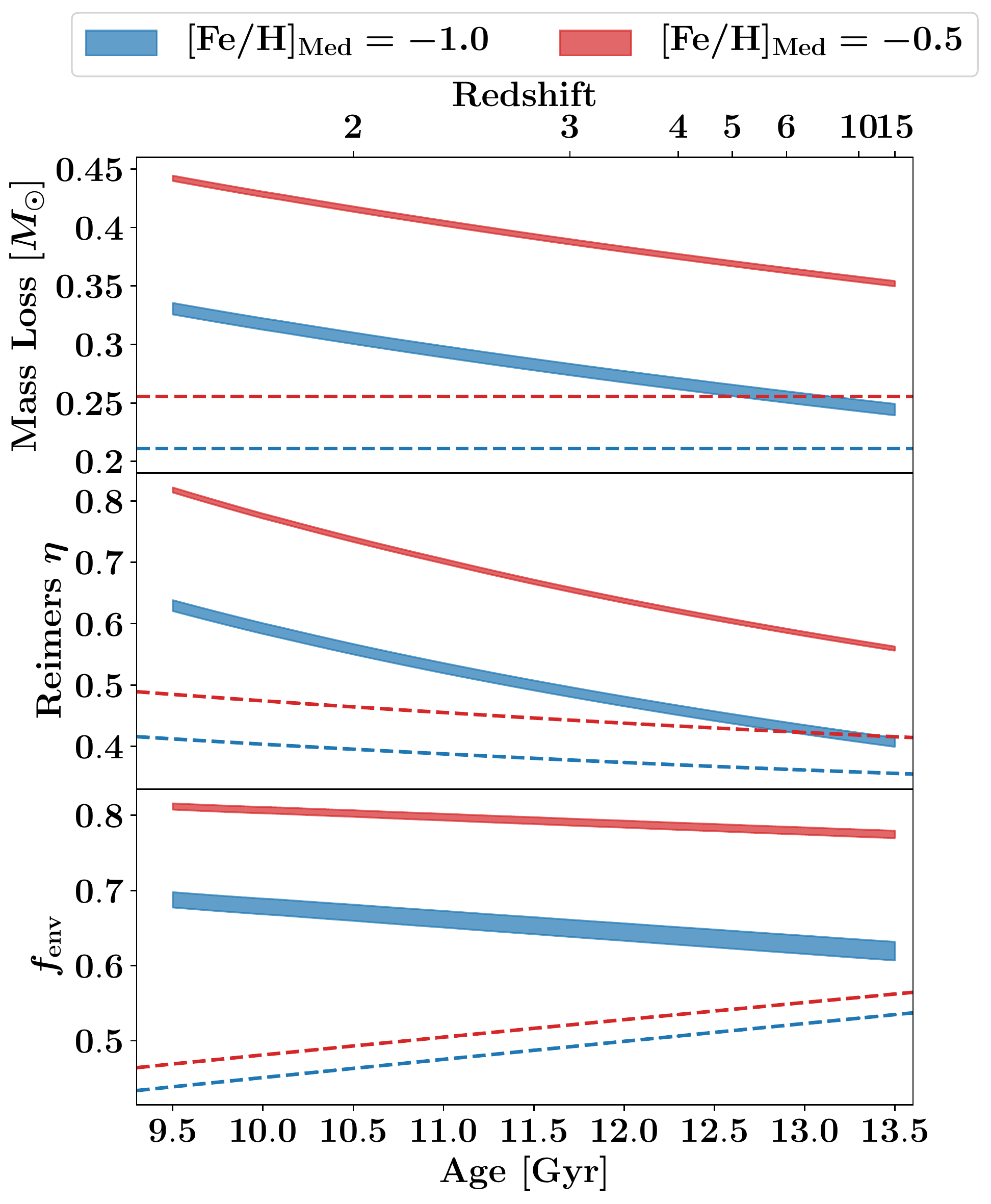}
\caption{Total red giant branch mass loss (upper), Reimers $\eta$ efficiency (middle) and fraction of stellar envelope lost (lower) required to obtain an RR Lyrae population with median [Fe/H] of $-1.0$ (blue band) and $-0.5$ (red band). The dashed line reports, for the same metallicities, the values obtained through the mass loss law of \citet{Savino19a}.}
\label{fig:requiredML}
\end{figure}

As we highlighted repeatedly in this paper, the RGB mass loss is a critical parameter in determining the properties of HB populations. Therefore, a first natural explanation for the existence of metal-rich RRLs is to postulate that these stars have lost a higher amount of mass during their RGB evolution than prescribed by S19. To provide a quantitative assessment of this scenario, we have generated a new set of stellar population models, increasing the zero point of the S19 prescription, until the median metallicity of the RRL population equals a given $[{\rm Fe/H}]_{\rm Med}$. We calculated the enhanced mass loss required to obtain $[{\rm Fe/H}]_{\rm Med}$ values of $-1.0$ and $-0.5$. The models have been generated using both the IS prescriptions of Sect.~\ref{Framework}. We used the Flat and Tail MDFs of Fig.~\ref{fig:MDFs}, as well as a Gaussian MDF analogous to the Peak models and centred on $[{\rm Fe/H}]_{\rm Med}$. The required mass loss, calculated at $[{\rm Fe/H}]_{\rm Med}$, is shown in Fig.~\ref{fig:requiredML}.  We found the variation introduced by the different stellar population models to be very small (see the width of the coloured bands).

A $[{\rm Fe/H}]_{\rm Med}$ of $-1.0$, corresponding to the upper boundary of our confidence interval for the peak of the bulge RRL population, can be obtained with a very old population and a mass loss value that is only slightly higher than the S19 prescription. This value rapidly increases for younger ages, reaching values above $0.33 M_{\odot}$ or Reimers $\eta > 0.6$ for the youngest models. Typical high values of $\eta$ are placed around 0.4-0.5 \citep{Catelan00,Jang14,McDonald15}, suggesting that, even accounting for a significantly higher mass loss than that of S19, the bulge RRL population is unlikely to have formed at redshifts lower than 4. The picture is even more extreme for $[{\rm Fe/H}]_{\rm Med} = -0.5$. The production of RRLs of such high metallicity requires RGB stars to lose $0.35-0.45 M_{\odot}$, or roughly 80\% of their hydrogen envelope mass.

Very high values of mass loss ($\eta \gtrsim 0.5$) are unlikely to be common among metal-rich RGBs, as that would be in contrast with observations of the local neighbourhood and of Galactic globular clusters \citep[e.g.][]{Layden95b,Salaris16,Tailo16}. However, it is still possible that extreme episodes of mass loss occur in a small fraction of stars. This has historically been suggested as an explanation for the existence of metal-rich RRLs \citep{Taam76, Layden95b}. In principle, such scenario would allow for the main RRL population of our sample to have much younger ages but, analogously to what seen in Sect.~\ref{Mass}, the consequent reduction in RRL production efficiency would be in strong conflict with the mass budget of the Galactic bulge. As for the tail of our sample close to solar metallicity, the smaller number of RRLs and the larger mass of the bulge stellar population at this metallicity mean that sporadic episodes of extreme mass loss are a valid hypothesis for the origin of these variable stars.

%----------------------------------------------------------------------------------------------------------------------------
%--------------------------------------      BINARIES      ------------------------------------------------------------
%----------------------------------------------------------------------------------------------------------------------------

\subsection{Binarity}
A possible solution to attain the high values of mass loss required for metal-rich RRLs lies in binary interaction.  As a star ascends on the RGB, the hydrogen envelope expands and becomes less gravitationally bound. The tidal field from a relatively close companion could further reduce the binding energy of the inflated and turbulent envelope, that is subsequently lost through stellar wind \citep{Eggleton89, Han02}. This scenario has the advantage of being more efficient in field stars than in globular clusters, given that clusters have low binary fractions \citep{Sollima07, Milone12b, Ji15}, so that the paucity of metal-rich RRLs in the latter is justified. Binarity in RRLs is very difficult to asses through standard radial velocity techniques, as the signal is blended with the much stronger radial velocity variation associated with pulsation, requiring long-term high precision spectroscopic campaigns \citep[e.g.][]{Guggenberger16}. A much more efficient approach makes use of the light-travel time effect \citep{Li-qian14,Hajdu15,Liska16}, which manifests as timing variations in the light-curve maximum of the RRL due to its position change as it orbits the companion.

\begin{figure}
\centering
\includegraphics[width=0.5\textwidth]{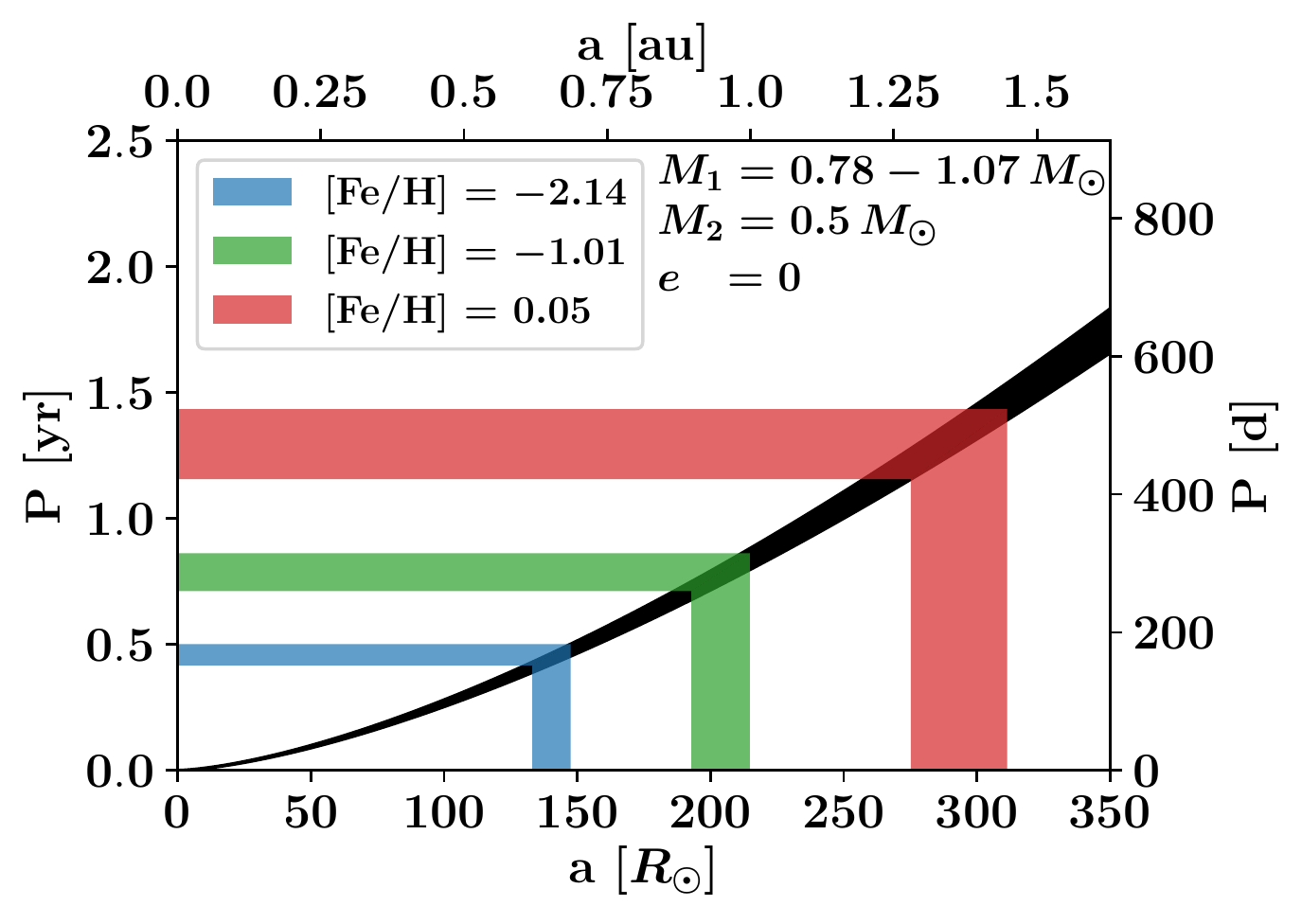}
\caption{Orbital period, as function of semi-major axis, for a circular binary system composed of an RGB primary and a $0.5 M_{\odot}$ secondary (black band). The shaded coloured bands mark, for different metallicities, the separation below which Roche lobe overflow occurs during the primary RGB evolution. The width of the bands is calculated adopting an age range of 9.5-13.5~Gyr.}
\label{fig:Roche}
\end{figure}

Therefore, we looked for potential binary systems in our sample by searching in the literature, and by utilizing photometric data from the OGLE database \citep{Soszynski14,Udalski15} to construct and analyse $O-C$ diagrams\footnote{Observed minus Calculated epoch of a given light-curve extreme (usually time of brightness maximum), using the stars' ephemerids. Pulsation period and times of brightness maximum determined by the OGLE survey \citep{Soszynski14}.} for our RRL sample, according to the prescriptions of \citet{Hajdu15} and \citet{Prudil19}. Searching for signs of binarity in photometric data of RRLs can be hampered by several effects that can mimic a binary in the $O-C$ diagram, such as the Blazhko effect \citep[quasi-periodic modulation of the light curve][]{Blazhko07}. In addition, the lack of long-baseline monitoring for some RRLs can greatly reduce detection efficiency. In general, half of our sample is affected by the aforementioned problems and was therefore excluded from this analysis. 

Among the remaining stars, a few dozens show semi-periodic behaviour in their $O-C$ diagrams, which should be further investigated, but only one robust binary candidate was found (OGLE-BLG-RRLYR-11522), therefore the binary-enhanced mass loss scenario is not strongly supported. The OGLE dataset limits our sensitivity to binaries with periods $P\lesssim 10 $~yr so, in principle, longer period systems are not excluded. However, such systems are likely too wide to experience significant interaction. The peculiar binary fraction of RRLs has been recently subject of discussion. While systems with period above several tens of years seem to be common \citep{Liska16, Li-qian18, Kervella19}, the binary fraction at smaller periods appears to be no more than a few percent \citep{Richmond11,Hajdu15,Prudil19}, with very few candidates  with $P \lesssim 7-10$~yr.

Of course one can expect a period threshold below which binary systems cannot produce RRLs. An RGB star with a very close companion will, in fact, eventually achieve Roche lobe overflow and initiate mass transfer. As the companion is statistically likely to have lower mass (either a white dwarf or a low-mass main sequence star), in most cases the envelope stripping will proceed rapidly until either a degenerate helium core or a hot subdwarf star are left \citep{Han02}. This critical period threshold depends on the properties of the binary system and on the metallicity of the RGB star. Fig.~\ref{fig:Roche} shows the period-separation relation for a binary system composed of an RGB star, in the age range 9.5-13.5~Gyr and for a range of metallicities, and a 0.5~$M_{\odot}$ companion, in a circular orbit. Using the stellar radius from the BaSTI models, and calculating the Roche lobe radius according to \citet{Leahy15}, we can infer the minimum period below which the primary fills its Roche lobe during the RGB ascent. This threshold is lower than 0.5~yr for metal-poor binaries and below 1.5~yr for solar-metallicity systems.

While this explains the absence of binaries with $P\lesssim 1.5$~yr, it does not justify the lack at $1.5\, {\rm yr}\lesssim P \lesssim 10$~yr. This tension can be alleviated by increasing the system eccentricity, so that the binary period increases but mass transfer can still be triggered at periastron. However, the eccentricities needed to achieve Roche lobe overflow in a 10~yr binary are very high, being at least 0.70 in solar metallicity pairs and at least 0.85 in metal-poor systems. We stress that the reasoning adopted here is quite idealised so, while it is sufficient to rule out the existence of very short period RRL binaries, a detailed  binary population synthesis approach should be taken to predict the orbital property distribution of RRL binaries in the 1-20 yr period interval and how well they match to the empirical evidence.

%----------------------------------------------------------------------------------------------------------------------------
%--------------------------------------      HELIUM      -----------------------------------------------------------------
%----------------------------------------------------------------------------------------------------------------------------

\subsection{Helium abundance}
\label{helium}
The last hypothesis concerns the chemical mixture of the stars in our sample. A first element of uncertainty is the value of $[\alpha/{\rm Fe}]$ of our RRLs. Stellar evolution is mostly sensitive to the total metal content, rather than iron abundance \citep{Salaris93}. Therefore decreasing the $[\alpha/{\rm Fe}]$ of our models would increase, roughly by the same amount, the resulting $[{\rm Fe/H}]_{\rm Med}$. However, the value of the  $[\alpha/{\rm Fe}]$ plateau in the bulge stellar population is known within 0.1~dex \citep[e.g.][]{Hill11, Friaca17, Bensby17}, so this effect is likely to be minor.

Far more important can be the impact of helium abundance. So far, we have assumed that the helium abundance steadily increases proportionally to metallicity. This is a standard practice in stellar evolution calculation \citep[e.g.][]{Bressan12,Choi16} and, in the case of BaSTI, the adopted relation is $\rm \Delta Y / \Delta Z = 1.4$, obtained by linear regression of the primordial helium abundance and the solar value \citep{Basti}. While reasonable, this assumption is not guaranteed to apply, as stars with anomalously high helium abundance have been observed to exist in, e.g., globular clusters \citep[e.g.][]{Piotto07,Dalessandro13,Marino14}. At fixed age and metallicity, the effect of increasing the abundance of helium is both a decrease of the stellar mass at the tip of the RGB and a decrease of atmospheric opacity in HB stellar models. These two effects combined move HB population models to significantly higher effective temperatures. The impact on the variable star population is that high-metallicity stars, that with a canonical chemical mixture would have not become RRLs, can enter the IS and manifest as metal-rich RRLs. 

\begin{figure}
\centering
\includegraphics[width=0.5\textwidth]{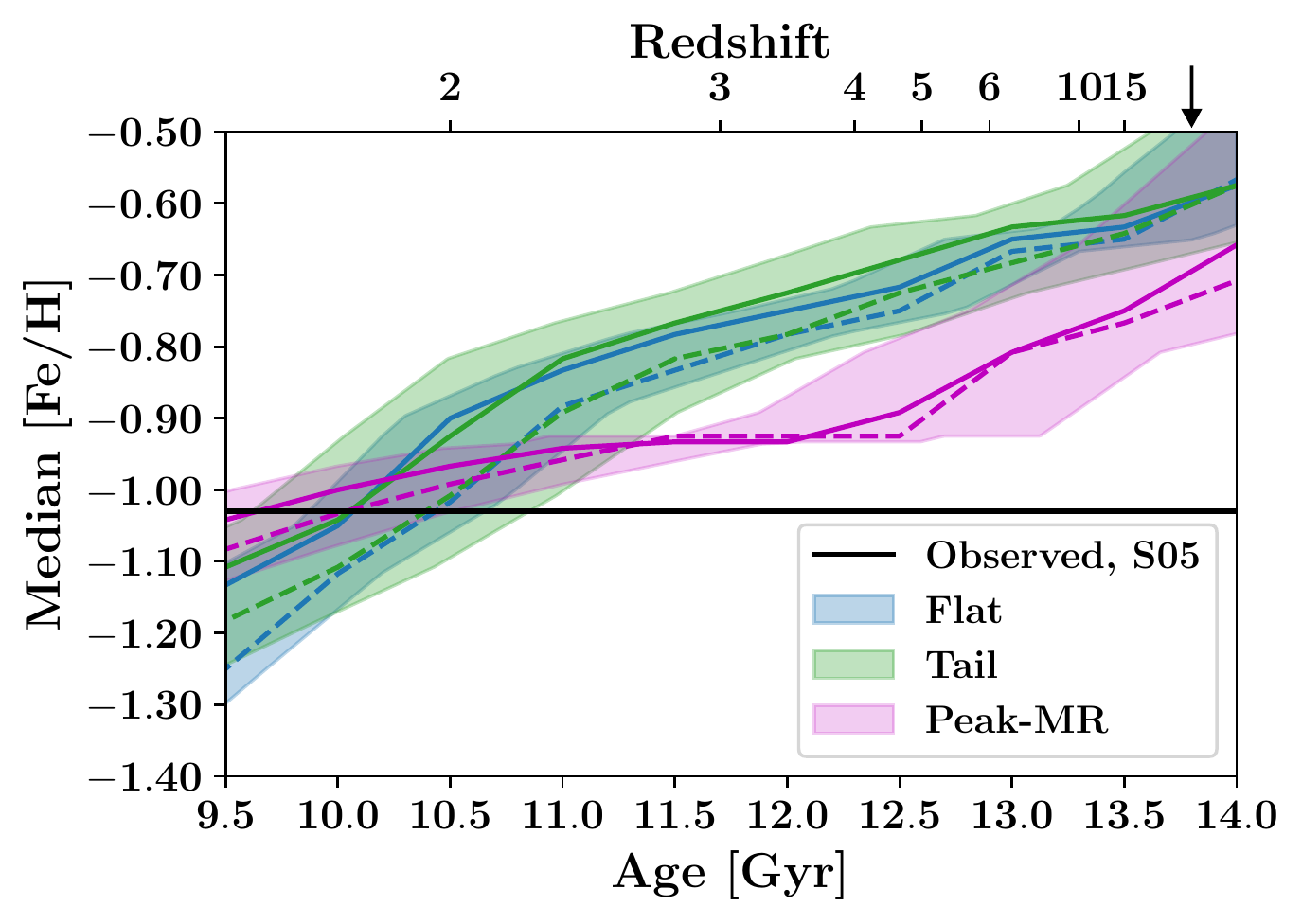}
\caption{Same comparison of Fig.~\ref{fig:modelhi}, but for stellar population models with helium mass fraction of Y~=~0.3.}
\label{fig:helium}
\end{figure}

We explore this scenario by recalculating our stellar population models with the BaSTI helium-enhanced isochrones and HB tracks \citep{Pietrinferni-he}. We adopted a helium abundance of Y~=~0.3. This value represents a moderate helium abundance enhancement, ranging from 0.03 to 0.055 depending on metallicity, and it is in line with the abundance spread observed in most Galactic globular clusters \citep{Milone18}. The outcome of these models is shown in Fig.~\ref{fig:helium}. In contrast to Fig.~\ref{fig:modelhi}, now RRL populations with $[{\rm Fe/H}]_{\rm Med} = -1.0$ can be produced even at ages lower than 10~Gyr, meaning that even a much more modest helium enhancement than Y~=~0.3 could reconcile our stellar population models with the RRL metallicity peak measured through the S05 method.

The hypothesis that the main bulge RRL population is helium enhanced is an intriguing possibility. The analysis of APOGEE data led \citet{Schiavon17} to the discovery of a sizeable population of RGB stars in the bulge with a light-element abundance pattern analogous to that of globular cluster stars. This sample of stars shows interesting similarities with the RRL population, in that it constitutes a few percent of the bulge total mass budget and it has a relatively narrow MDF peaking around $[{\rm Fe/H}] = -1.0$. If, as observed in globular clusters, this peculiar light-element pattern is associated with an enhancement of a few per cent in helium abundance, then the helium-burning progeny of these RGB stars could manifest as an RRL population compatible in number and metallicity with that observed in the bulge. The possible enhanced helium abundance in bulge RRLs was already suggested by \citet{Lee16} to explain their period properties. However, the study of \citet{Marconi18} argue, by means of hydrodynamical pulsation models, that the helium abundance of the main RRL population is compatible with the cosmological value but does not exclude the presence of a minor population with small helium enhancements.

With respect to the metal-rich tail of the RRL population, Fig.~\ref{fig:helium} shows that Y$=0.3$ is not sufficient to account fo their existence within a Hubble time, requiring even higher helium mass fractions to produce RRLs with $[{\rm Fe/H}] \gtrsim -0.5$. Such high Y values are rather extreme and only observed in the most massive globular clusters \citep{Milone18}. We note, however, that the most metal-rich RRLs were not included in the analysis of \citet{Marconi18}, so we have no empirical evidence to refute such hypothesis. On the whole, the issue of helium abundance in both the main population and the metal-rich tail of bulge RRLs should be settled through spectroscopic investigation. Helium abundance is notoriously difficult to measure spectroscopically and, while a detection could still be achieved in the hottest phases of RRL pulsation, we argue that a much more robust and efficient approach would be to look for light-element abundance anomalies, under the assumption that they are connected to helium as they are in globular clusters. Sodium abundance, in particular, is not affected by evolutionary effects or binary interactions during the RGB phase and would constitute a reliable tracer of globular cluster-like helium enhancements.

%----------------------------------------------------------------------------------------------------------------------------
%----------------------------------------------------------------------------------------------------------------------------
%--------------------------------------      CONCLUSIONS      ------------------------------------------------------
%----------------------------------------------------------------------------------------------------------------------------
%----------------------------------------------------------------------------------------------------------------------------

\section{Discussion and conclusions}
\label{Conclusions}
In this paper we attained precision dating of the central stellar spheroid of the Milky Way by considering the properties of RRLs in the bulge within a stellar population framework. We derived CaT-based metallicities for 935 stars in the BRAVA-RR dataset, obtaining the first spectroscopic MDF for a large sample of RRLs in the bulge. We compared the spectroscopic measurements with photometrically derived metallicities and considered the uncertainties related to different calibrations and metallicity scales to identify a 0.35~dex confidence interval for the MDF peak of the main RRL population.

We built a set of synthetic stellar population models and concluded that the high metallicity of the RRL population is an indication of an extremely ancient age, the oldest determined to date among the Milky Way stellar populations. Albeit with significant error bars, we determine a formation redshift of roughly 10-15. Our age precision excludes formation redshifts lower than 6, although this limit can be lowered to 4 by relaxing the assumptions in our population models. We stress that this age measurement does not represent a mean age for the inner spheroidal stellar population. RRLs, and the BHBs used for analogous studies in the stellar halo, represent in fact a biased age tracer, as they are only sensitive to the oldest stellar populations and the presence of a younger population that produces more massive helium-burning stars cannot be ruled out by this analysis (although the size of this young spheroidal population is strongly constrained by the bulge total mass budget). This measurement should therefore be interpreted as the mean age at which significant formation of population II stars began in what is now the inner stellar spheroid.

The age measured here does not represent the formation time of the spheroidal structure in the Milky Way centre, either. Given the hierarchical paradigm of galaxy formation \citep{White78}, it is in fact likely that this ancient star formation took place in smaller galactic progenitors that later reassembled into the stellar structure we observe today in the centre of the Galaxy. This scenario would in fact be compatible with theoretical models showing that, while at redshifts between 10 and 15 the first proto-Milky Way fragments were still in the early phases of coalescence, they were already experiencing intense star formation and starting to reionise the local environment \citep[e.g.][]{Li14, Aubert18}, compatible with the earliest known records of starburst activity in the high-redshift Universe \citep[e.g.][]{Hashimoto18,Lam19}.

We stress that, in contrast with the general notion that time and metallicity are closely related, it is not unreasonable to have such an ancient and metal-rich stellar population. The link between stellar population age and metallicity is very sensitive to to the underlying star formation history and the chemical clock speeds up in dense and violent star formation environments. Numerical simulations show that high-redshift stellar populations can enrich to $\sim0.1\,Z_{\odot}$ within a few hundreds million years, provided they reside in haloes that experience intense merging, and star formation, histories \citep[e.g.][]{Wise12, Salvaterra13, Yajima15}. These characteristics are likely to apply to the Milky Way proto-bulge and indeed chemical evolution models show that a large fraction of the bulge likely enriched to slightly sub-solar metallicities within $\sim 0.5$~Gyr \citep[e.g.][]{Grieco12, Matteucci19}.

This scenario is further supported by the fact that some of the oldest globular clusters in the Galaxy (such as NGC6522, NGC6558 and HP1) reside in the bulge and have metallicities of $[{\rm Fe/H}] \sim -1.0$ \citep[e.g.][]{Barbuy09,Ortolani11,Barbuy18b,Kerber19}. The properties of old globular clusters in the bulge are a nice and independent confirmation of the results presented in this manuscript. It has been long suggested that some of the massive bulge clusters, even significantly more metal-rich than $[{\rm Fe/H}] \sim -1.0$ \citep[e.g. Terzan 5,][]{Massari14}, could be relics of the early stages of bulge formation. The results of this paper show that field stars in the bulge do not need to be as metal-poor as the halo to be equally old, or even older.

The age value inferred in this paper for the Milky Way central RRLs agrees with extrapolation of the old population gradient found in the stellar halo, once differences in the stellar population analysis are taken into account. The comparison with the innermost stellar halo age determinations, and the absence of any metallicity gradient in our RRL sample, seems to confirm the presence of the ``ancient chronographic sphere" reported by \citet{Carollo16}, where the age gradient flattens. The presence of a transition in the stellar spheroid properties at $r_{GC}\sim 5$~kpc is a possibility that should be further investigated through additional chemical and kinematical tracers.

As our stellar population framework does not account for the production of RRLs with $[{\rm Fe/H}] \gtrsim -1.0$, we also explored possible formation channels for metal-rich RRLs. While we conclude that sporadic episodes of high RGB mass loss are a valid channel to produce a small population of very metal-rich RRLs, it seems that the only way to produce the the main RRL population at much younger ages than reported in this paper is to allow an  increase of a few per cent in the helium mass fraction. A conclusive answer on this latter possibility has yet to be established but, while its refutation would strengthen the case for a truly ancient RRL population, its confirmation would open to equally interesting implications for the formation of the central stellar spheroid. A helium-enhanced RRL population would naturally suggest a connection with the N-rich sample of \citet{Schiavon17} and with the formation scenarios the authors propose for their stars, namely that they were stripped from a population of globular clusters, that they share some of the formation mechanisms with globular clusters or, a scenario that would confirm the results of this paper, that they carry the chemical imprint of the first stellar generations \citep[e.g.][]{Meynet10,Chiappini11}.

As several uncertainties still surround the properties and nature of the Milky Way central RRLs, we conclude outlining some directions for future improvements to help settle the remaining issues. As the age of RRL populations so steeply depend on metallicity, an updated calibration of spectroscopic and photometric indicators for these stars, based on large samples, as well as robust conversions between metallicity scales, would help reducing the uncertainty on where the peak of the population is truly located. A revised calibration of CaT-based metallicities is especially encouraged, as it would be of great value for the large sample of Galactic RRL spectra that \textit{Gaia} will provide. A precise metallicity for bulge RRLs can also be obtained by means of high-resolution spectroscopy. Beside providing a validation of less direct methods, this approach will have the advantage of enabling studies of the detailed abundance pattern of the RRL population, including helium-tracing light elements like sodium. As the age-metallicity calibration presented here critically depends on the RGB mass loss, additional investigation is urged, on larger and more diverse samples of stellar systems, to increase measurement robustness and quantify variations with age metallicity and environment. Regarding the topic of metal-rich RRLs, a full binary population synthesis framework needs to be explored so that the precise contribution of mass transfer and other binary interactions can be established. Finally, we anticipate that the large spectroscopic surveys of the coming decade, like 4MOST \citep{deJong12}, MOONS \citep{Cirasuolo14} and SDSS-V \citep{Kollmeier17} will not only help address some of the issues raised above but will also provide a much more complete characterisation of the metal-poor tail of the bulge MDF. The plethora of data these surveys will provide, coupled with improved stellar models and increasingly reliable dating for field stellar populations, will allow us to exploit multiple and complementary archaeological indicators, and gain precious insights into the early formation phases of the heart of our Galaxy.

\begin{acknowledgements}
We thank the anonymous referee for helping us improve this manuscript. A.S. thanks R.M. Rich, C. Wegg, L. Posti and G. Iorio for insightful discussions at various stages of this manuscript preparation. A.S., A.K. and Z.P. acknowledge support by the Deutsche Forschungsgemeinschaft (DFG, German Research Foundation) -- Project-ID 138713538 -- SFB 881 (``The Milky Way System", subproject A11). Z.P. acknowledges the support of the Hector Felllow Academy. R.S. was supported by the National Science Center, Poland, Sonata BIS project 2018/30/E/ST9/00598. This research made use of the following Python packages: \texttt{Astropy} \citep{Astropy12}, \texttt{IPython} \citep{IPython}, \texttt{Matplotlib} \citep{Matplotlib}, \texttt{NumPy} \citep{Numpy}, \texttt{pandas} \citep{Pandas} and \texttt{SciPy}\citep{Scipy}.
\end{acknowledgements}

\bibliographystyle{aa}
\bibliography{./Bibliography}

\end{document}